\documentclass[a4paper,notitlepage]{article}

\usepackage[top=2cm,bottom=2cm,left=2cm,right=2cm,includefoot]{geometry}
\usepackage{setspace}
\usepackage{indentfirst}
\usepackage[margin=10pt,font=small,labelfont=bf,labelsep=quad,figurename=Figure.,tablename=Table.]{caption}

\usepackage{ccaption}

\usepackage{multirow}	
\usepackage{array}
\usepackage{threeparttable}
\usepackage{color, colortbl}
\usepackage{makecell}

\definecolor{Sandy}{rgb}{0.93, 0.92, 0.88}
\definecolor{LightCyan}{rgb}{0.88,1,1}
\definecolor{Gray}{gray}{0.9}

\usepackage[utf8]{inputenc}
\usepackage{newtxtext,newtxmath}
\usepackage{bm} 
\usepackage{amsmath, amstext, amsfonts}
\usepackage[T1]{fontenc} 
\usepackage{times} 
\usepackage[scaled]{helvet}

\usepackage[numbers,super,sort&compress]{natbib}
\usepackage{mciteplus}
\makeatletter
\renewcommand\@biblabel[1]{#1.}
\makeatother

\usepackage{graphicx}

\usepackage{hyperref}
\hypersetup{colorlinks,citecolor=green,linkcolor=blue}

\usepackage[pagewise]{lineno}

\usepackage[version=4]{mhchem} 
\usepackage[ruled]{algorithm2e} 

\usepackage{xr}

\usepackage{authblk} 

\usepackage{textcomp}

\begin{document}

\title{Self-organized learning emerges from coherent coupling of critical neurons}
\author[a]{Chuanbo Liu}
\author[b, $\dagger$]{Jin Wang}
\affil[a]{State Key Laboratory of Electroanalytical Chemistry, Changchun Institute of Applied Chemistry, Chinese Academy of Sciences, Changchun, Jilin, 130022, P.R. China.}
\affil[b]{Department of Chemistry and of Physics and Astronomy, State University of New York, Stony Brook, 11794-3400, New York, USA.}
\date{\today \\ $^\dagger$ jin.wang.1@stonybrook.edu}
\maketitle

\begin{abstract}
	Deep artificial neural networks have surpassed human-level performance across a diverse array of complex learning tasks, establishing themselves as indispensable tools in both social applications and scientific research.  
	Despite these advances, the underlying mechanisms of training in artificial neural networks remain elusive.  
	Here, we propose that artificial neural networks function as adaptive, self-organizing information processing systems in which training is mediated by the coherent coupling of strongly activated, task-specific critical neurons.  
	We demonstrate that such neuronal coupling gives rise to Hebbian-like neural correlation graphs, which undergo a dynamic, second-order connectivity phase transition during the initial stages of training.  
	Concurrently, the connection weights among critical neurons are consistently reinforced while being simultaneously redistributed in a stochastic manner.  
	As a result, a precise balance of neuronal contributions is established, inducing a local concentration within the random loss landscape which provides theoretical explanation for generalization capacity.  
	We further identify a later on convergence phase transition characterized by a phase boundary in hyperparameter space, driven by the nonequilibrium probability flux through weight space.
	The critical computational graphs resulting from coherent coupling also decode the predictive rules learned by artificial neural networks, drawing analogies to avalanche-like dynamics observed in biological neural circuits.  
	Our findings suggest that the coherent coupling of critical neurons and the ensuing local concentration within the loss landscapes may represent universal learning mechanisms shared by both artificial and biological neural computation.
\end{abstract}

\section*{Introduction}

Artificial neural networks (ANNs), especially the deep ones, have become the new silver bullet for almost all problems in scientific and social domains.
Examples including 
game playing \cite{silverMasteringGameGo2017}, 
language processing \cite{openaiGPT4TechnicalReport2023}, 
protein folding \& binding \cite{jumperHighlyAccurateProtein2021}, 
drug design \cite{zhangArtificialIntelligenceDrug2025}, 
material design \cite{merchantScalingDeepLearning2023}, 
math problem solving \cite{chervonyiGoldmedalistPerformanceSolving2025}, 
algorithm design \cite{novikovAlphaEvolveCodingAgent2025}, 
weather precast \cite{priceProbabilisticWeatherForecasting2025}, 
nuclear fusion \cite{seoAvoidingFusionPlasma2024}, 
and solving dynamics of complex systems \cite{liuDistillingDynamicalKnowledge2024}, etc. 
Numerous attempts had been made to develop a theory of how neural networks learn, from the perspectives of 
function optimization \cite{rumelhartLearningRepresentationsBackpropagating1986}, 
statistical physics \cite{hopfieldNeuralNetworksPhysical1982, bahriStatisticalMechanicsDeep2020}, 
predictive coding \cite{songInferringNeuralActivity2024}, 
neuroevolution \cite{stanleyDesigningNeuralNetworks2019}, 
Hebbian plasticity \cite{kuriscakBiologicalContextHebb2015}, 
information theory \cite{tishbyInformationBottleneckMethod2000}, 
and statistical and Bayesian learning \cite{vapnikNatureStatisticalLearning2000, fristonFreeenergyPrincipleUnified2010}. 

However, there remain foundational open questions about how ANN learns and predicts. 
For example, why can stochastic gradient descent (SGD) find a converged state in a highly rugged loss landscape, analogous to the Levinthal paradox in protein folding \cite{karplusLevinthalParadoxYesterday1997}? 
Moreover, this convergence is largely independent of diverse initializations, variations in hyperparameters, and different architectural choices, hinting at an inherent stability in the underlying training dynamics.
The predictive power of ANNs is also a mystery: why can ANNs adaptively master various tasks with the same model architecture, and why does their predictive ability generalize even to outliers in the training dataset? 
How can we interpret the knowledge learned by ANNs?
Finally, does training of ANNs share any similarity with biological neural circuits?

\section*{Results}

\setcounter{section}{0}
\setcounter{subsection}{0}
\section{Population dynamics of deep artificial neural network learning}
\subsection{Emergence of self-organized neuronal correlation structure during training}
\begin{figure}
	\centering
	\includegraphics[width=\linewidth]{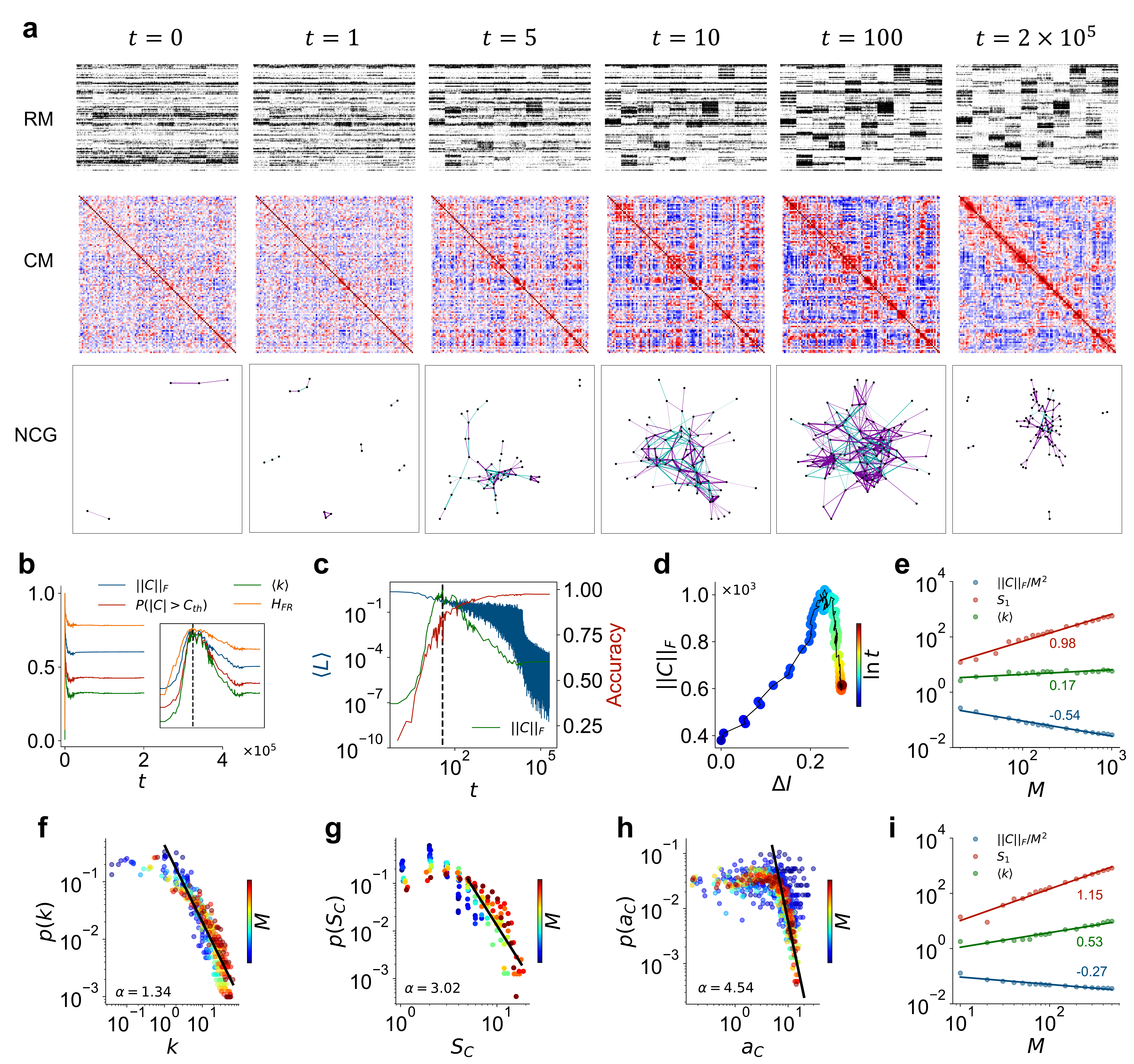}
	\medskip
	\caption{\textbf{Emergence of activity patterns and connectivity phase transition.}
	\textbf{a}, Visualization of representative Rastermap (RM), correlation matrix (CM), and neuronal correlation graph (NCG) of neuronal activities for training iterations $t=0,1,5,10,100,2\times 10^5$. Neurons are reordered based on the activities of the terminal model at $t=2\times 10^5$. The color range of each CM is $-1 \sim 1$. $C_{th} = 0.5$ is used to construct the NCG, and non-connected neurons are omitted. Purple and green lines in the NCG indicate positive and negative correlations, respectively. Line widths in each NCG are proportional to the correlation strength between neurons.
	\textbf{b}, Changes in the Frobenius norm $||C||_F$, the survival probability $P(|C| > C_{th})$ for $C_{th} = 0.5$, the mean degree $\left\langle k \right\rangle$, and the Forman-Ricci entropy $H_{FR}$ during the early stage of training. The vertical dashed line indicates the peak of $||C||_F$ at $t=37$.
	\textbf{c}, Comparison of $||C||_F$ with the oscillation of ensemble-averaged loss $\bar{L}$ and accuracy increment. The vertical dashed line is the same as in (\textbf{c}).
	\textbf{d}, Relationship between information gain $\Delta I$ and total correlation strength $||C||_F$.
	\textbf{e}, Dependency of the normalized Frobenius norm of the correlation matrix $||C||_F/M^2$, size of the largest connected component $S_1$, and the mean degree $\langle k \rangle$, for various model sizes $M$, at iteration $t=100$. 
	\textbf{f-h}, Dependency of degree distribution $p(k)$ (\textbf{f}), clique size distribution $p(S_C)$ (\textbf{g}), and distribution of clique-wise overall absolute neuronal activities $p(a_C)$ (\textbf{h}) on various model sizes $M = 30 \sim 1010$, at terminal iteration $t=2 \times 10^5$. Black lines indicates fitting all observations to a power-law distribution $p(x) \sim x^{-\alpha}$. 
	\textbf{i}, Dependency of $||C||_F/M^2$, $S_1$ and $\langle k \rangle$, for various model sizes $M$, at terminal iteration $t=2 \times 10^5$.
	}
	\label{fig:activity_Rastermap}
\end{figure}

To address these questions, we begin by examining the emergent collective dynamics of neuronal activity throughout the training process of ANNs.
We employ Rastermap \cite{stringerRastermapDiscoveryMethod2025}, an activity pattern discovery method from neuroscience, to uncover possible activity patterns formed during the training of a 3-layer fully-connected ANN (Methods). 
As shown in Fig.~\ref{fig:activity_Rastermap}a, persistent activity patterns are established for each learning task (digits $0 \sim 9$) early in training and remain largely unchanged until the final iteration. 
We confirm that this behavior is universal across different initialization methods, SGD hyperparameters, and model sizes (total number of neurons). 
The emergence of these activity patterns indicates the development of specific neuron correlation patterns, as demonstrated by the Pearson correlation coefficient matrix in Fig.~\ref{fig:activity_Rastermap}a. 
Interpreting the correlation coefficient matrix as an adjacency matrix yields a network which is analogous to the brain's functional connectome, enabling the application of standard network analytic techniques. 
In particular, thresholding produces a binarized graph, which we termed the neuronal correlation graph (NCG), as illustrated in Fig.~\ref{fig:activity_Rastermap}a. 
Notably, the evolution of neuronal activities exhibits a flourish-diminish process in both correlation strength and topology of NCG: neurons rapidly become correlated within a few ($\sim 10$) training iterations, forming an extensive correlated network ($\sim 100$), which then diminished to a smaller but more compact network by the final iteration ($\sim 2\times 10^5$).

The global correlation strength, quantified by the Frobenius norm of the correlation matrix $||C||_F$, exhibits a sharp shift during the early stage of training ($t \approx 0 \sim 100$) (Fig.~\ref{fig:activity_Rastermap}b). 
We designate the end of this shift $t = 100$ as the transition point.
This abrupt shift is also accompanied by an behavior divergence of the training loss: before the transition point, the loss decreases smoothly and rapidly; after the transition point, the loss decreases slowly with pronounced fluctuations (Fig.~\ref{fig:activity_Rastermap}c).

With the correlation threshold $C_{th}$ held fixed during training, modulating pairwise neuronal correlations produces distinct NCG topologies, reminiscent of classical percolation phenomena. 
To track these topological changes, we quantified the survival probability $P(|C| > C_{th})$, the mean degree $\left\langle k \right\rangle$, and the Forman-Ricci entropy $H_{FR}$ across training iterations. 
Each metric captures complementary aspects of NCG organization and collectively indexes the emergence of network structure. 
All three exhibit sharp, near-synchronous peaks that align with the transition in correlation strength (Fig.~\ref{fig:activity_Rastermap}b). 
Notably, this topological transition persists across thresholds spanning $C_{th} = 0.5 \sim 1.0$, indicating a global, rank-preserving shift of the network structure rather than a thresholding artifact. 
Prior to training, network organization is weak. 
As training proceeds, both the total number of edges and the size of the largest connected component undergo a flourish-diminish process, consistent with a transient, system-wide reorganization of correlation structure. 
This behavior is also evident for the initially random correlation matrix, computed from an initially randomly sampled batch of data: the edge-count distribution $p(N_E)$ displays elevated mean and variance near the transition point.

Alongside the evolution of strength and structural transitions, prediction accuracy increases rapidly prior to the transition point and then continues to rise more gradually thereafter (Fig.~\ref{fig:activity_Rastermap}c). 
To probe the sustained improvement in accuracy, we computed the prediction entropy from the last-layer activity distribution, $S_{pred}(t) = \left\langle - \ln p(a, t) \right\rangle$, which varies monotonically over training. 
The prediction entropy quantifies uncertainty in the ANN's outputs and enables a definition of information gain, $\Delta I(t) = S_{pred}(0) - S_{pred}(t)$. 
In parallel, the alignment of neuronal correlation strength and structure, enabling $||C||_F$ to serve as a proxy for computational complexity (in the Kolmogorov sense) under the assumption that computation is predominantly mediated by highly correlated neurons, as broader neuronal participation implies a longer effective program (i.e., more numerical operations). 
Using $||C||_F$ and $\Delta I$, we identify a distinct turning point in the complexity-information phase diagram (Fig.~\ref{fig:activity_Rastermap}d), with computational complexity first increasing and then receding, while information gain grows steadily. 
The terminal stage of training maps to a fixed point in this diagram. 
Moreover, the cumulative fraction of explained variance continuous to increase for leading principal components, indicating a progressive preference for reducing representation dimensionality of data-target pairs, consistent with the evolution of the eigenvalue spectra. 
A sparsity metric, defined as the exponent obtained by fitting the eigenvalue spectra of correlation matrix to a power law \cite{stringerHighdimensionalGeometryPopulation2019}, exhibits a zigzag trajectory over training, coincident with the reversal observed in the complexity-information diagram and aligned with the flourish-diminish dynamics of neuronal connectivity.

\subsection{Second-order connectivity phase transition in early training stage}

To interrogate the mechanistic basis of the flourish-diminish dynamics in neuronal connectivity and the concomitant reversal of network computational complexity, we quantified topology transitions in the NCG using a suite of graph-theoretic measures. 
Across all metrics, we observed a unified and sharply delineated flourish-diminish transition, indicating a coherent reconfiguration of NCG connectivity during training. 
To assess scale dependence, we examined the dependency of these measures to hyperparameters. 
At the transition, the measures exhibit clear signature of scale-divergence behavior, with $||C||_F \sim M^{3/2}$, $S_1 \sim M$, and $\left\langle k \right\rangle \sim 1$, where $M$ is the number of nodes in the NCG (i.e. the model size) (Fig.~\ref{fig:activity_Rastermap}e). 
It turns out the transition is insensitive to batch-size variation but shifts to earlier iterations under larger learning rates. 
A pronounced jump in the susceptibility $\chi$ marks the emergence of a giant component, analogous to the percolation transition in Erd\"{o}s-R\'enyi networks.
However, connectivity phase transition gradually diminishes for larger models, suggesting that increasing $M$ allows the model to enhance its complexity by utilizing more neurons for feedforward computations, thereby forming extensive correlated graphs in the final state. 

In the context of complex networks \cite{dorogovtsevCriticalPhenomenaComplex2008}, the size of the largest connected component plays the role of a correlation length in thermodynamic systems; its scaling as \(\mathcal{O}(M)\) in finite systems signals divergence. 
Consistently, the susceptibility peaks and diverges at the transition. 
From an intensity viewpoint, the abrupt change in $||C||_F$ indicates a concomitant shift across all thresholds $C_{th}$ (Fig.~\ref{fig:activity_Rastermap}b). 
Taken together, these signatures identify a continuous (second-order) phase transition: the NCG evolves smoothly, while the derivatives of the order parameters ($||C||_F$, $S_1$, $P(|C| > C_{th})$, $H_{FR}$, etc) exhibit discontinuities. 
We refer to this flourish-diminish (complexity-information) transition as \emph{connectivity phase transition}, underscoring its origin in training-induced changes to neuronal connectivity. 
Notably, this transition does not involve barrier crossing; the emergence of connected components and cliques reflects collective neuronal dynamics rather than rare-event jumps.

\subsection{Scale-free correlation structure in final training stage}

At the transition point of connectivity phase transition, neurons are extensively correlated and susceptibility is elevated, consistent with a highly connected architecture at criticality. 
By contrast, in the terminal training phase of small models ($M \le 400$; MNIST; three-layer fully connected ANN), the NCG exhibits reduced connectivity with fragmented components (Fig.~\ref{fig:activity_Rastermap}b), indicating a departure toward a percolation-like regime. 
Nevertheless, large models exhibit a muted connectivity phase transition, with the trained model attaining the highest level of connectivity. 

Despite these divergent connectivity regimes, at the final iteration the degree distribution displays finite-range power-law pattern across all sizes of models (Fig.~\ref{fig:activity_Rastermap}f). 
The power-law distribution indicates that when the scale is changed, the distribution does not change its shape. 
Therefore, the distribution is scale invariant. 
Strength-independent scale-invariant behavior likewise characterizes component sizes, clique sizes (Fig.~\ref{fig:activity_Rastermap}g), and their activity-valued counterparts (sum of neuronal activities within a component or clique) (Fig.~\ref{fig:activity_Rastermap}h). 
Across correlation thresholds, these observables are well described by power laws with finite-size cutoffs (Fig.~\ref{fig:activity_Rastermap}f-h). 

These findings imply that the trained model resembles a scale-free correlation structure, with a small number of high-degree hubs dominate connectivity through hierarchical organization of neurons.  
The scale-averaged representative activity patterns, as well as the fraction of strongly correlated neuron pairs, are expected to remain invariant with increasing network size. 
In such networks, hubs create ultra-short paths that compress communication latency and accelerate global coordination, while preserving sparsity that lowers wiring and maintenance costs. 
Consequently, scale-free correlation structures in trained models enhance computational and functional capacity via efficient communication, a core requirement for ANN functionality. 
The power-law distribution of clique sizes further indicates that the NCG is strongly heterogeneous, comprising dense cores populated by large cliques and a broad periphery containing only small cliques. 
Networks with core-periphery or overlapping-community organization, in which some groups are highly cohesive, are robust to random failures yet vulnerable to targeted perturbations of core nodes or inter-community bridges. 
Such robustness to random connectivity fluctuations is expected in ANNs subjected to random feedforward operations (e.g., dropout) and data sample variance, whereas targeted disruption of core neuronal correlation can precipitate failures in adversarial attack and continual learning \cite{finlaysonAdversarialAttacksMedical2019, dohareLossPlasticityDeep2024}.

Notably, the fitted exponent for the degree distribution is approximately $1.34$, slightly below the typical value ($2 \sim 3$) for natural scale-free networks \cite{choromanskiScaleFreeGraphPreferential2013}, yet consistent with the distribution at the connectivity transition. 
In contrast, the clique-size exponent in the terminal model, $\sim 3.02$, substantially exceeds the value $\sim 1.97$ observed at the connectivity transition point. 
The terminal model further exhibits scale-divergent behavior, with $||C||_F \sim M^{7/4}$, $S_1 \sim M$, and $\left\langle k \right\rangle \sim M^{1/2}$, which deviates modestly from the scale dependencies at the connectivity transition point (Fig.~\ref{fig:activity_Rastermap}i). 
These observations indicate that the terminal model retains a form of topological criticality, albeit with less extensive neuronal correlations. 
Accordingly, we may characterize the NCG in the terminal regime as quasi-critical: critical features are principally adhered, but slightly derivates from a perfect criticality at the connectivity transition point \cite{tianTheoreticalFoundationsStudying2022}. 

\setcounter{section}{1}
\setcounter{subsection}{0}
\section{Measure concentration and flatness of loss landscape}
\subsection{Local concentration of loss landscapes}
\begin{figure}
	\centering
	\includegraphics[width=\linewidth]{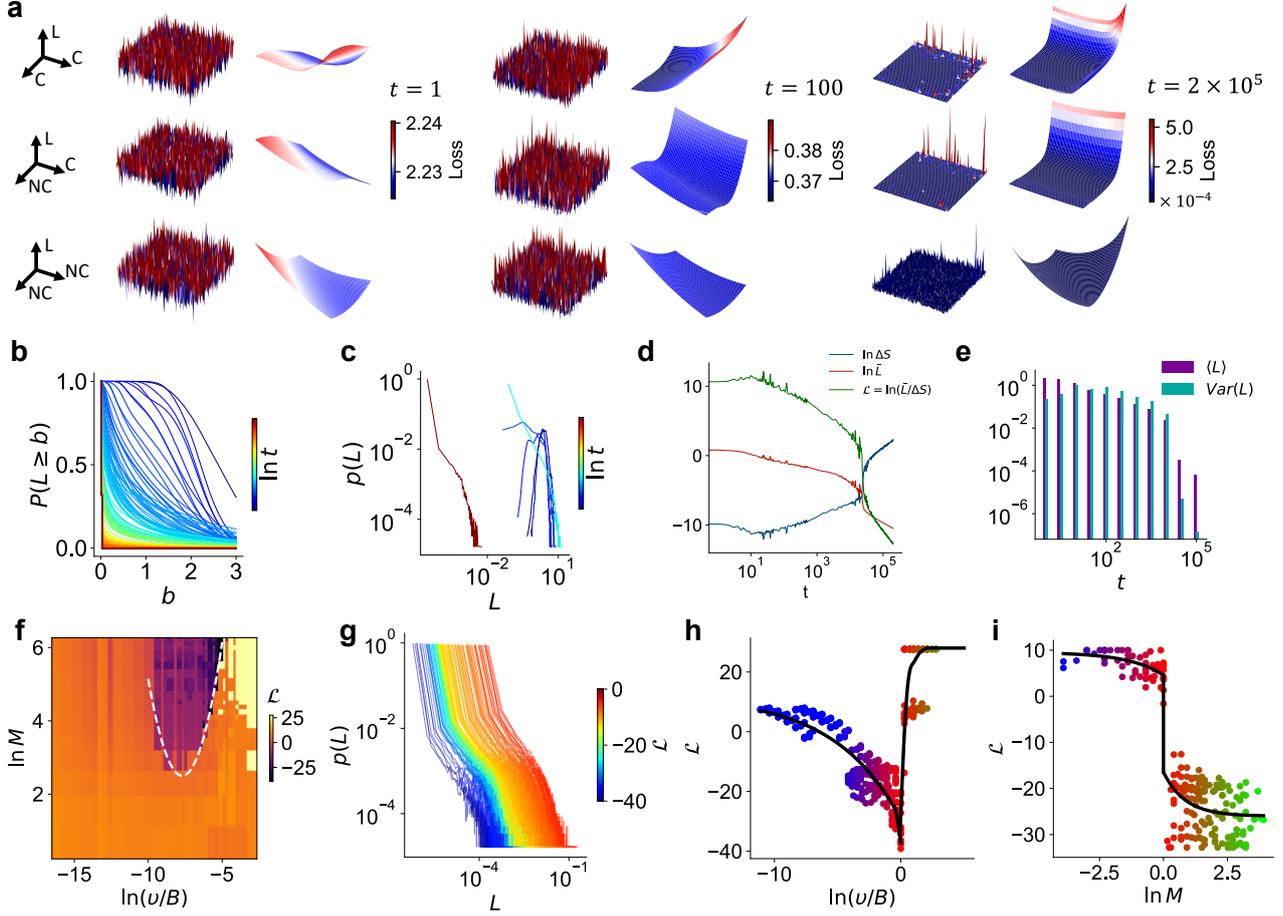}
	\medskip
	\caption{\textbf{Local concentration of loss landscapes and convergence phase transition.}
	\textbf{a}, Visualization of representative effective ($\hat{L}$, $B=64$, left) and mean landscapes ($\bar{L}$, right) for critical (C) and non-critical (NC) connections. Iterations $t=1, 100, 2 \times 10^5$ are shown for comparison.  
	\textbf{b}, Change in concentration inequality $P(L \ge b)$ as a function of threshold value $b$ across training iterations $t$.  
	\textbf{c}, Distribution of per-sample landscapes $p(L)$ with respect to training iteration $t$.  
	\textbf{d}, Change in entropy difference $\Delta S = \ln N - S$, average loss $\bar{L}$ and convergence criterion  $\mathcal{L} = \ln (\bar{L}/\Delta S)$.  
	All entropy computations use a large deviation parameter $b = 5 \times 10^{-3}$. Entropy differences for $b > \max \{L\}$ are extrapolated using spline fitting.
	\textbf{e}, Change of landscape mean $\bar{L}$ and variance $Var(L)$ with learning iteration.
	\textbf{f}, The phase diagram of training is determined by the convergence criterion $\mathcal{L}$, model size $M$, and the mobility factor $\gamma = \upsilon / B$.  
	The empirical boundary of the converged phase is depicted as a white dashed line, defined by the function $\varphi(\gamma) = \frac{1}{2} (\gamma + 7.7)^2 + 2.5$.
	\textbf{g}, Distribution of per-sample loss $p(L)$, for various hyperparameter configurations satisfy $\ln \frac{M}{\varphi(\gamma)} > 1$ and $\mathcal{L} < -5$.
	\textbf{h-i}, Convergence criterion $\mathcal{L}$ for varying mobility factor $\gamma = \upsilon / B$ (\textbf{h}), and model size $M$ (\textbf{i}). For each $\gamma$ ($M$) varys,  $\mathcal{L}$ is collected with $M$ ($\gamma$) held constant. For (\textbf{h}), x-axis is shifted to align the minimum values of each collection to 0; for (\textbf{i}), x-axis is shifted to align the point of steepest gradient of each collection to 0.  
	}
	\label{fig:landscape}
\end{figure}

Upon observing dynamic alterations in NCG topology throughout the training process, we sought to elucidate the corresponding changes in the loss landscape.
From the perspective of the loss landscape, SGD navigates the training process via the averaged landscape determined by a batch of randomly sampled data, with each iteration employing a new data batch. 
We refer to this site-dependent, batch-averaged landscape as the empirical loss landscape, denoted by $\hat{L}(w)$ for the connection weight vector $w \in \mathbb{R}^{d_w}$. 
Since the averaged content changes randomly for every weight vector in the parameter space, $\hat{L}$ constitutes a random field whose statistical properties are directly related to those of the loss landscapes. 
As expected for a random field, $\hat{L}$ is highly rugged almost everywhere, in contrast to the much smoother ensemble-averaged landscape $\bar{L}$ over all data samples (Fig.~\ref{fig:landscape}a, left and right panels). 
Fundamentally, the roughness of $\hat{L}$ arises from the divergence of per-sample landscapes $L = \{L_x\}$ ($x \in \mathcal{D}$, where $\mathcal{D}$ is the dataset with $N$ samples), while batch averaging effectively smooths out this divergence. 
Indeed, since the covariance of $\hat{L}$ is inversely proportional to batch size $B$, and $\bar{L}$ corresponds to setting $B=N$, $\bar{L}$ represents the most smoothed version of $\hat{L}$. 
As training progresses, both the roughness of $\hat{L}$ and the value of $\bar{L}$ decrease, eventually reaching a state where the roughness nearly disappears and $\bar{L} \to 0$ (Fig.~\ref{fig:landscape}a, right panel). Consequently, $\bar{L}$ becomes flat near the converged state.

As shown in Fig.~\ref{fig:landscape}b, the alteration of the concentration probability $P(L \ge b)$ for a given threshold $b$ indicates that per-sample losses gradually concentrate, but the probability of high losses drops sharply within a few training iterations. 
Notably, the form of the per-sample distribution undergoes a dramatic shift from a log-normal to a power-law-like heavy-tailed distribution when crossing the transition point (Fig.~\ref{fig:landscape}c). 
Moreover, we observe that the decrements in the roughness of $\hat{L}$ and in $\bar{L}$ do not occur simultaneously (Fig.~\ref{fig:landscape}e). 
Before the transition point at $t \sim 100$, the roughness increases, while $\bar{L}$ decreases significantly from above 2.22 to below 0.4. 
After the transition point, the roughness then reduces substantially. 
Generally, when the per-sample loss becomes heavy-tailed, both the mean and variance of the empirical landscape are dominated by exceptionally large loss data, i.e., rare events \cite{vezzaniSinglebigjumpPrinciplePhysical2019}. 
Thus, it is clear that the training process first reduces the per-sample loss for the majority of the population before the transition point, and then primarily focuses on a small set of exceptional data with large losses. 

Surprisingly, this concentration of measure phenomenon is not confined to a specific point. 
Suppose the weight space of an ANN exhibits certain invariant transformations, such as permutation~\cite{pmlr-v139-simsek21a}, rescaling~\cite{10.5555/3305381.3305487,10.5555/3045118.3045167}, rotation~\cite{ziyin2022on}, or data-weight duality~\cite{fengActivityWeightDuality2023}. 
It can be demonstrated that these invariant transformations preserve the statistical properties of the empirical landscape. 
If the loss landscape is concentrated at $w^*$, then every point connected to $w^*$ via an invariant transformation will exhibit analogous concentration.
Alternatively, assuming that the per-sample loss landscape $L$ varies only minimally within a small local basin, concentration at the basin center guarantees concentration across the entire basin. 
Consequently, a locally concentrated subspace emerges in which both the mean and variance of the empirical landscape approach zero. 
The vanishing variance further indicates that divergence among batched data samples is suppressed, thereby reducing the noise strength of SGD and yielding diminished entries in the gradient covariance matrix. 
Because the gradient covariance and the Hessian are aligned in the neighborhood of $w^*$~\cite{yangStochasticGradientDescent2023}, reduced gradient covariance concomitantly signals reduced curvature of the averaged loss landscape, thereby resulting in a flat local subspace.
If the samples in the test dataset can be mapped to those in the training dataset via invariant transformations (possibly due to shared key features), it follows that the loss landscape for the test data is also concentrated at $w^*$. 
Consequently, the prediction accuracy on the test dataset will match that obtained on the training dataset, indicating that the neural network generalizes effectively.

We refer to the behavior that all empirical loss landscapes concentrate and approach a degenerate distribution near the same optimal solution within a local subspace as the \emph{local concentration of loss landscapes}. 
Remarkably, as an atypical property of ANNs, for any meaningful learning task, a converged model yields uniform performance: every data sample incurs a low error, rather than only achieving a low ensemble average. 
Note a subtle difference between local uniformity of random fields and local concentration of loss landscapes: local uniformity only requires that all random fields have constant statistical properties (mean, variance, covariance structure, etc.) within a sufficiently small neighborhood, whereas local concentration further requires that all loss landscapes concentrate to the same minimum loss in a local region with the same (obviously constant) statistical properties. 

\subsection{Quasi-thermal free energy and convergence criterion}

The concentration of empirical loss landscapes admits a large-deviation bound:
\begin{equation}\label{eq:entropy_variational_principle}
	P(\hat{L} \geq b) \leq e^{-B\Delta S},
\end{equation}
where $b$ denotes a rare-event threshold, $B$ is the batch size, and $\Delta S = \ln N - S \equiv \left\langle -\ln (1/N) \right\rangle - \left\langle -\ln p_\beta \right\rangle$ measures the information gap between the quasi-thermal Gibbs distribution $p_\beta = \frac{1}{Z} e^{-\beta L}$ and the uniform distribution over the training dataset $\mathcal{D}$ of size $N$. 
We refer to $\Delta S$ as the quasi-thermal entropy (hereafter, entropy), which is exactly the large-deviation rate function.
Large $\Delta S$ ensures exponentially rapid collapse of the landscape distribution to its ensemble average $\bar{L}$; hence, the monotonic increase of $\Delta S$ during training indicates a maximum entropy-like principle. 
The functionality of the ANN requires that $\bar{L}$ attains a theoretical minimum value $L_0$. 
To balance the simultaneous minimization of $\bar{L}$ and maximization of $\Delta S$, we propose a minimum free energy principle:
\begin{equation}\label{eq:principle}
	w^* = \inf_w \mathcal{F} := \inf_w \left\langle E \right\rangle - \lambda S
\end{equation}
where $\mathcal{F}$ is the information free energy, $\left\langle E \right\rangle = \bar{L}$ quantifies the model's performance, $S$ measures the upper bound on performance divergence, and $\lambda \le 0$ is a hyperparameter controlling their relative contributions. 
The negativity of $\lambda$ follows from the fact that loss function are lower bounded, thereby right bound on the tail of $L$ implies $-\beta > 0$, which in turn yields $\lambda \le 0$. 
$\mathcal{F}$ serves as a Lyapunov function when both the majority population and the exceptional large-loss data samples exhibit negative gradients, which is consistent with the behavior of the loss landscape distribution during training (Fig.~\ref{fig:landscape}b,c). 
Subsequently, the training of ANNs is a process that aims to optimize for the worst case across all data samples.
For local concentration of loss landscapes, the consistent statistical properties of loss landscapes within the local region indicates $\mathcal{F}$ will also be minimized within this region. 
Thus, $\mathcal{F}$ also provides a robust estimate of the generalization capability of ANNs.

The introduction of thermodynamic-like concepts also enable the understanding of loss landscape behaviors. 
As indicated in Fig.~\ref{fig:landscape}d, a clear sudden shift of $\Delta S$ can be identified at $t\approx 4\times 10^4$, suggesting the shift of ``driving forces''. 
Indeed, obvious turnaround occurs for the gradients of $\Delta S$ and $\bar{L}$. 
The first stage (before $t\approx 4\times 10^4$) is dominated by changes in $\bar{L}$, while the second stage is mainly determined by changes in entropy. 
Consider $\bar{L}$ is analogous to ``energy'', we can term the first stage as ``energy-driven'', and the second stage as ``entropy-driven''. 
This is consistent with the expansion range of the entropy distribution, suggesting that attempts to minimize $\bar{L}$ result in a decrease of $\Delta S$, consequently increases the upper bound of the loss and leads to a more divergent empirical loss landscape.
Similar behaviors are observed for variance and average per-sample loss, where we have observed $Var(L) \ge \epsilon(t) \left\langle L \right\rangle^2, \forall t$, with $\epsilon(t)$ exhibits a turnaround as $t$ increases. 

In practice, since the size of trainng dataset $\mathcal{D}$ is finite, we instead compute the equivalent convergence criterion $\mathcal{L} = \ln (\bar{L}/\Delta S)$ to quantify learning convergence. 
As shown in Fig.~\ref{fig:landscape}d, after several initial training iterations, $\mathcal{L}$ consistently decreases until convergence is achieved. 
It is important to note that both $\Delta S$ and $\bar{L}$ are macro-variables of a single stochastic trajectory and thus experience fluctuations during training.

\subsection{First-order convergence phase transition in final training stage}
The qualification of convergence enables us to evaluate the impacts of hyperparameters, such as the initial condition, learning rate $\upsilon$, minibatch size $B$, and model size $M$, on training dynamics and the performance of the converged model under various hyperparameter settings.

We find that the behavior of the effective loss $\hat{L}$ and test accuracy is similar for $\upsilon$ and $B$, suggesting that the training divergence caused by these two factors can be captured by a single variable. 
Therefore, we introduce the mobility factor $\gamma = \upsilon / B$ by analyzing the continuous-time stochastic dynamics of training using SGD, which provides a unified phase diagram for $\upsilon$ and $B$.
Note that test accuracy is a relatively coarse metric of assessing training, as it depends on the argmax over class probabilities, thereby ignores the distribution of confidence assigned to non-target classes.  
The effective loss $\hat{L}$ is a random variable and thus cannot provide a deterministic criterion for training behavior. 

On the contrary, both $\bar{L}$ and $\Delta S$ offer a fine-grained description of the training behavior. 
However, it should be noted that $\bar{L}$ quantifies the model's functionality, while $\Delta S$ measures the upper bound of performance divergence. 
The logarithm ratio $\mathcal{L} = \ln (\bar{L} / \Delta S)$ provides a unified quantification of training behavior, as illustrated in the phase diagram of Fig.~\ref{fig:landscape}f. 
A clear phase boundary, $\ln M = \varphi(\ln \gamma)$ where $\varphi$ is a parabola function, appears as a U-shape in the $M$-$\gamma$ hyperparameter space. 
Since model size quantifies the computational cost of feedforward inference, the U-shaped phase boundary implies a unique optimum, which is converged and of minimal size, for training a fully connected three-layer ANN on the MNIST dataset.

For large $\gamma$, $\bar{L}$ diverges and $\Delta S \to 0$, resulting in gradient explosion. 
Conversely, for small $\gamma$, $\bar{L}$ remains large and $\Delta S$ remains small. 
A small $\mathcal{L}$ corresponds to heavy-tailed per-sample loss landscapes (Fig.~\ref{fig:landscape}g).
Convergence is achieved only for small $\mathcal{L}$, hence $\mathcal{L}$ serves as a reliable indicator of training convergence. 
The form of the per-sample loss landscape distribution is also sensitive to the relative position of the phase boundary $\varphi(\ln \gamma)$. 
When far below the phase boundary, e.g., $\ln M / \varphi(\gamma) < -2$, the per-sample loss distribution approaches a log-normal distribution.
Crossing this convergence phase boundary corresponds to a first-order phase transition, as indicated by the order parameter $\mathcal{L}$ (Fig.~\ref{fig:landscape}h,i). 
Since an abrupt performance change occurs across the phase boundary, we term the first-order transition from the unconverged to the converged state as the \emph{convergence phase transition}.

\subsection{Cooperation of hyperparameters determines training fate}

Beyond the static phase diagram, the macro-observables of the training process reveal additional information about learning. 
The distance from the phase boundary, $\ln M / \varphi(\gamma)$, serves as an indicator of the quality of the training hyperparameters. 
And the behavior of $\Delta S$ elucidates the divergence of the loss distribution: a small $\Delta S$ implies a loose upper bound on the right tail of the loss function, indicating that the initial loss landscape is diverged. 
For large positive distances and diverged initial loss landscapes, training is characterized by monotonically increasing entropy and decreasing average loss. 
In contrast, for concentrated initial loss landscapes, the ANN first diverges its loss landscapes, driven by the average loss until $t\sim 10^4$, and then entropy increases slowly. 
For diverged initial conditions but suboptimal hyperparameters, corresponding to a very small learning rate or very large batch size, the convergence process is very slow but still exhibits monotonically decreasing $\mathcal{L}$, indicating that given enough training iterations, these models can still converge. 
Therefore, with sufficiently good hyperparameters, training can almost always converge, regardless of whether the initial loss landscapes is diverged or concentrated.

However, if the hyperparameters are not suitable, i.e., $\ln M / \varphi(\gamma)$ has a large negative value, training convergence is not guaranteed. 
For the diverged initial loss landscapes, the average loss decreases monotonically, but the increase in entropy is insufficient to achieve final convergence within $2\times 10^5$ training iterations. 
Even with sufficient training iterations, these training trajectories remain too noisy to ultimately converge. 
The situation is worse for the concentrated initial loss landscapes. 
Although the hyperparameters are better, the average loss abruptly increases with a sharp decrease in entropy, resulting in highly fluctuating average loss and entropy, indicating that the final stage will never converge. 
The worst case occurs when both the initial loss landscapes are concentrated, and the learning hyperparameters are poor. 
Even though the average loss still decreases monotonically, the entropy change decreases and fluctuates hopelessly, leading to an increase in $\mathcal{L}$.


\subsection{Understanding phase transition and generalization with probability flux}

\begin{figure}
	\centering
	\includegraphics[width=\linewidth]{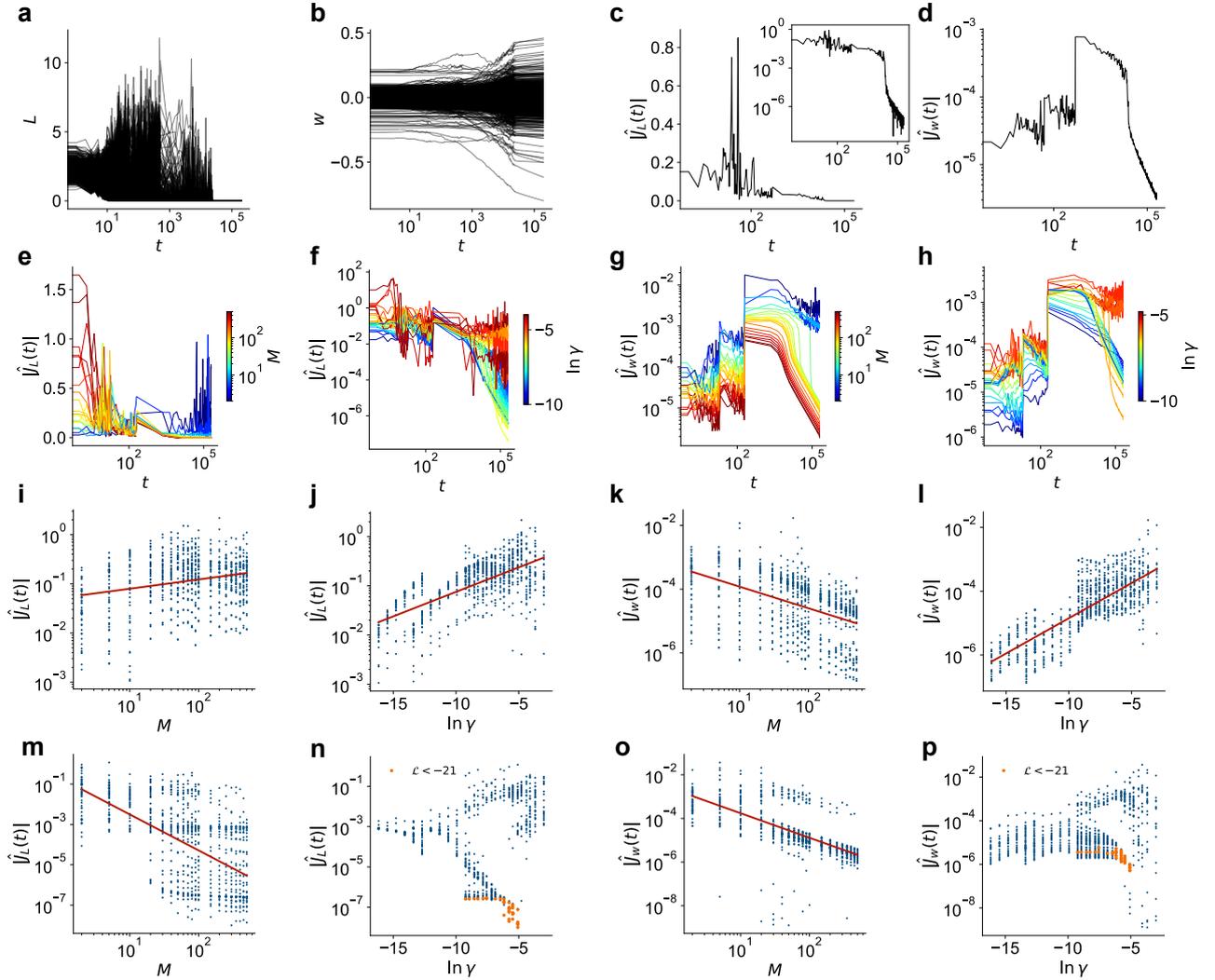}
	\medskip
	\caption{\textbf{Marginal probability flux signature phase transitions of training.}
	\textbf{a}, Marginal trajectory ensemble of the time-dependent per-sample loss $L(t)$.  
	\textbf{b}, Marginal trajectory ensemble of the time-dependent connection weights $w_i(t)$.
	For (\textbf{a}) and (\textbf{b}), 1,000 randomly sampled trajectories are shown for illustration purposes. 
	\textbf{c, d}, Dependence of the absolute total marginal probability flux of per-sample loss $|\hat{J}_L(t)|$ (\textbf{c}) and connection weights $|\hat{J}_w(t)|$ (\textbf{d}), on the training iteration $t$. 
	\textbf{e--h}, Evolution of $|\hat{J}_L(t)|$ (\textbf{e, f}) and $|\hat{J}_w(t)|$ (\textbf{g, h}) during training as functions of the model size $M$ (\textbf{e, g}) and mobility factor $\gamma$ (\textbf{f, h}), respectively. 
	\textbf{i--l}, Fitting of the relations $\ln |\hat{J}_L(t)| \propto \ln M$ (\textbf{i}), $\ln |\hat{J}_L(t)| \propto \ln \gamma$ (\textbf{j}), $\ln |\hat{J}_w(t)| \propto \ln M$ (\textbf{k}), and $\ln |\hat{J}_w(t)| \propto \ln \gamma$ (\textbf{l}) at iteration $t = 20$, respectively. 
	\textbf{m--p}, Fitting of the relations $\ln |\hat{J}_L(t)| \propto \ln M$ (\textbf{m}), $\ln |\hat{J}_L(t)| \propto \ln \gamma$ (\textbf{n}), $\ln |\hat{J}_w(t)| \propto \ln M$ (\textbf{o}), and $\ln |\hat{J}_w(t)| \propto \ln \gamma$ (\textbf{p}) at iteration $t = 2 \times 10^5$, respectively. 
	Each dot in (\textbf{i--p}) represents an individual training instance with different hyperparameter values.
	}
	\label{fig:flux}
\end{figure}

To identify a single macro-variable that accounts for both connectivity and convergence phase transitions while mitigating the curse of dimensionality, we analyze the marginal dynamics of per-sample loss and connection weight at the transition point across alternative dynamical regimes (Fig.~\ref{fig:flux}a,b). 
The trajectories are strongly stochastic and are well captured by a generalized Langevin dynamics
\begin{equation}\label{eq:generalized_langevin}
	\frac{d w}{dt} = - g(w, t) + \eta(w, t) 
\end{equation}
where $g = - \frac{\partial \bar{L}}{\partial w}$ denotes the gradient of the averaged loss landscape, $\eta$ is the site-dependent time-varying noise with covariance $\left\langle \eta(w, t) \eta^T(w, t') \right\rangle = 2D \delta(t-t')$. $D$ is the diffusion tensor characterizing the strength and anisotropy of the fluctuations, $D = \frac{\upsilon}{2B} H$ near the converged state $w^*$. 

Since the trajectory of $w(t)$ is stochastic, one can not predict the precise location of $w(t)$. 
However, the statistical pattern or probability evolves under the constraints of local conservation of probability, and thus can be predicted using the Fokker-Planck equation 
\begin{equation}\label{eq:Langevin_fokker_planck}
	\frac{\partial p(w, t)}{\partial t} = - \nabla J, \qquad J(w, t) = g(w, t)p(w, t) - \nabla (Dp(w, t))
\end{equation}
The physical meaning is clear. 
The evolution of probability is determined by the flux $J(w, t)$ in or out of each site. 
This probability flux is from deterministic driving force as the gradient of the loss function $g$, and the fluctuation or diffusion related contribution. 
At steady-state, we expect $\frac{\partial p_{SS}(w, t)}{\partial t} = 0$, thus $\nabla J_{SS} = 0$. 
If $J_{SS} = 0$, this means net flux in or out of any sites are all zeros. 
This is the equilibrium situation with detailed balance. 
On the other hand, if net flux in or out of the system is not zero, then $\nabla J_{SS} = 0$ implies that $J_{SS}$ is rotational since there is no sink or source for the flux to go into or come out from. 
The nonzero flux $J_{ss}$ is a measure of the degree of detailed balance breaking or noequilibrium.
Therefore, the driving force can be decomposed of three components \cite{fangNonequilibriumPhysicsBiology2019a, wangLandscapeFluxTheory2015a, wangPotentialLandscapeFlux2008, wangQuantifyingWaddingtonLandscape2011}
\begin{equation}\label{eq:driving_force_jin}
	g = - D \cdot \nabla U + \frac{J_{SS}}{p_{SS}} + \nabla \cdot D 
\end{equation}
where $U = - \ln p_{SS}$. 
In other words, the driving force for ANN training is determined by the gradient of the steady-state probability or potential landscape $- D \cdot \nabla U$, the steady-state probability flux force $\frac{J_{SS}}{p_{SS}}$, and the derivative of fluctuation characterized by diffusion $\nabla \cdot D$. 
The gradient component, \(-D \cdot \nabla U(w)\) with \(U=-\ln p\), draws the system toward high-probability attractor states and thus stabilizes the prevailing attractor. 
By contrast, the steady-state probability flux \(J_{\mathrm{SS}}\) is rotational, and tends to drive the system to go around without being limited to the point attractor. 
whereas the time-dependent flux encodes transition probabilities between states.
Finally, the divergence of the diffusion tensor, \(\nabla \cdot D(w,t)\), reshapes and shifts the landscape, modulating the position and robustness of attractor states.

It is noted that the driving force can also be decomposed to the time-dependent components \cite{wuPotentialFluxField2014}
\begin{equation}\label{eq:driving_force_jin_time_dependent}
	g = - D \cdot \nabla U(w, t) + \frac{J(w, t)}{p(w, t)} + \nabla \cdot D(w, t), \quad U(w, t) = - \ln p(w, t)
\end{equation}
In summary the driving force of SGD for ANN training is determined by the gradient of the time-dependent probability landscape $- D \cdot \nabla U(w, t)$, the time-dependent probability flux $\frac{J(w, t)}{p(w, t)}$, and derivative of the diffusion with respect to connection weight $\nabla \cdot D(w, t)$. 
Note the same decomposition can be applied to any other random elements that constitute a stochastic process, e.g. the per-sample loss. 

The probability flux of the associated Fokker-Planck equation is expected to delocalize probability mass and tends to destabilize local attractors in favor of inter-state transitions. 
Consequently, the flux provides the nonequilibrium driving necessary to dismantle an existing attractor and nucleate or select alternative attractors, thereby mediating the phase transition. 
For example, increasing flux can induce a saddle-node bifurcation that converts a single-attractor regime into a bi-stable one, whereas decreasing flux can reverse this process and restore monostability. 
The probability flux as the driving force for nonequilibrium phase transition/bifurcation has been shown for several complex and biological system \cite{liUncoveringUnderlyingMechanisms2020, suDynamicalThermodynamicalOrigins2024, wangEarlyWarningIndicators2024, wangUnderstandingUnderlyingPhysical2024, wenboUncoveringUnderlyingMechanism2017, xuCurlFluxDynamical2020, xuNonequilibriumEarlywarningSignals2023a, yanThermodynamicDynamicalPredictions2023, zhangConnectionDissipativeChaos2023, zhangDynamicThermodynamicOrigin2020}. 
It is likewise anticipated that flux governs phase transitions in the training dynamics of ANN. 
Accordingly, we expect a pronounced change in flux near the connectivity transition, as substantial flux variations destabilize existing attractor states and, in extreme cases, generate or reshape a comparatively flat landscape that facilitates optimal flux propagation.

Since the flux is a multidimensional vector, in order to avoid computational complexity, we explore the marginal flux to represent the effect of flux by reducing significant computational burden. 
We propose the marginal loss flux, defined as $J(L, t) = \frac{1}{N} \sum_k \dot{L}_k(t) \delta (L - L_k(t))$ for per-sample loss $L_k$ of $x_k \in \mathcal{D}$ for training dataset $\mathcal{D}$ of size $N$, and the marginal weight flux $J(w, t) = \frac{1}{N} \sum_i \dot{w}_i(t) \delta (w - w_i(t))$ for single-variable connection weight $w_i$. 
The total marginal fluxes are then defined as $|\hat{J}_L(t)| = \int |J(L, t)| dL$ and $|\hat{J}_w(t)| = \int |J(w, t)| dw$, respectively, which quantify the total velocity of probability density shift for loss and weight distributions. 
If there exists significant probability flux from one state to another, this signals an increased likelihood of a phase transition. 
At or near the critical point, fluxes become enhanced, indicating a higher rate of spontaneous phase transitions. 
Therefore, an abrupt increase in $|\hat{J}_L(t)|$ and $|\hat{J}_w(t)|$ serves as a signature of phase transitions in the probability distributions of per-sample loss and connection weight, respectively.

As illustrated in Fig.~\ref{fig:flux}c,d, the substantial shift in the $|\hat{J}_L(t)|$ and $|\hat{J}_w(t)|$ signals indicates the connectivity phase transition during training. 
The transition point occurs at approximately $t \sim 100$ and $t \sim 10^4$. 
The former corresponds to the connectivity transition point, as evident from graph topological measures of NCG (Fig.~\ref{fig:activity_Rastermap}b). 
The later corresponds to the convergence trainsition as evident from convergence criterion, and the shift of driving forces from average loss to entropy. 

For the early training stage ($t = 0 \sim 100$), increasing model size $M$ results in an increase in $|\hat{J}_L(t)|$ (Fig.~\ref{fig:flux}e,i) and a decrease in $|\hat{J}_w(t)|$ (Fig.~\ref{fig:flux}g,k). 
Consequently, a large $M$ can achieve sufficiently large per-sample loss distribution changes with limited connection weight distribution changes. 
As the result, an increase in $M$ is associated with a decrease in $|\hat{J}_w(t)|$ for adequate $|\hat{J}_L(t)|$ , thereby slowing the weight exploration dynamics and resulting in delayed or diminished phase transition (Fig.~\ref{fig:activity_Rastermap}e). 
For learning rate $\upsilon$ and batch size $B$, variations in these hyperparameters produce similar effects and can be uniformly quantified by the mobility factor $\gamma$ (Fig.~\ref{fig:flux}f,h). 
An increase in $\gamma$ consistently leads to increased total marginal probability flux in the early stage of training (Fig.~\ref{fig:flux}f,h,j,l). 
These behaviors explain the shift in connectivity transition point: large $\gamma$ indicates large $|\hat{J}_w(t)|$, leading to faster exploration of the weight space and larger $|\hat{J}_L(t)|$, consequently resulting in earlier connectivity phase transition. 

In the final stage of training ($t = 10^4 \sim 2 \times 10^5$), quite the opposite occurs: large model size decreases both $|\hat{J}_L(t)|$ and $|\hat{J}_w(t)|$ (Fig.~\ref{fig:flux}m,o).
Meanwhile, increasing the mobility factor leads to a bifurcation in $|\hat{J}_L(t)|$ and $|\hat{J}_w(t)|$ for the final stage of training, with the branch of smaller total flux corresponding to converged models (Fig.~\ref{fig:flux}n,p). 
Therefore, both larger model size and higher mobility factor (given sufficiently large model size) promote a decrease in total marginal probability flux, suggesting that the final stage has reached a stationary basin with minimal probability current. 

This divergence in behavior is caused by a shift in driving forces: the early stage of training is primarily driven by average loss, while the late stage is mainly driven by entropy. 
In the initial stage, larger models act like more powerful machines that output more ``energy'', thus accelerating the phase transition. 
The mobility factor acts as the inverse mass that transforms force into velocity. 
In the final stage, larger model size indicates stronger representational power, enabling the model to find a more concentrated loss landscapes. 
Conversely, a higher mobility factor indicates a greater ability to explore the weight space, which is essential for locating a wide concentrated region of loss landscape, resulting in even smaller flux in the terminal state. 

This also holds for the complete probability flux near the converged state with concentrated loss landscapes. 
The flux in this concentrated region is exponentially related to the mobility factor $\gamma$ and barrier height $\Delta \bar{L}$, i.e. $J \propto e^{-\Delta \bar{L}/\gamma}$. 
Thereby, increasing (decreasing) learning rate $\upsilon$ or reducing (increasing) batch size $B$ will significantly increase (decrease) the flux. 
Conversely, lower (higher) barrier height in loss landscape will significantly be associated with higher (lower) flux. 
However, excessively high mobility will cause divergence in the training dynamics, as seen in the phase diagram Fig.~\ref{fig:landscape}f.

Overall, phase transitions in ANN training can be consistently attributed to variations in probability flux across both the loss landscape and the connection-weight space. 
Elevated flux accelerates the onset of connectivity phase transition and may improve generalization by steering optimization toward broader, locally concentrated regions of the loss landscape.

\setcounter{section}{2}
\setcounter{subsection}{0}
\section{Emergence of criticality in deep artificial neural network learning}
\subsection{Heavy-tailed neuronal connectivity and criticality in training}

\begin{figure}
	\centering
	\includegraphics[width=\linewidth]{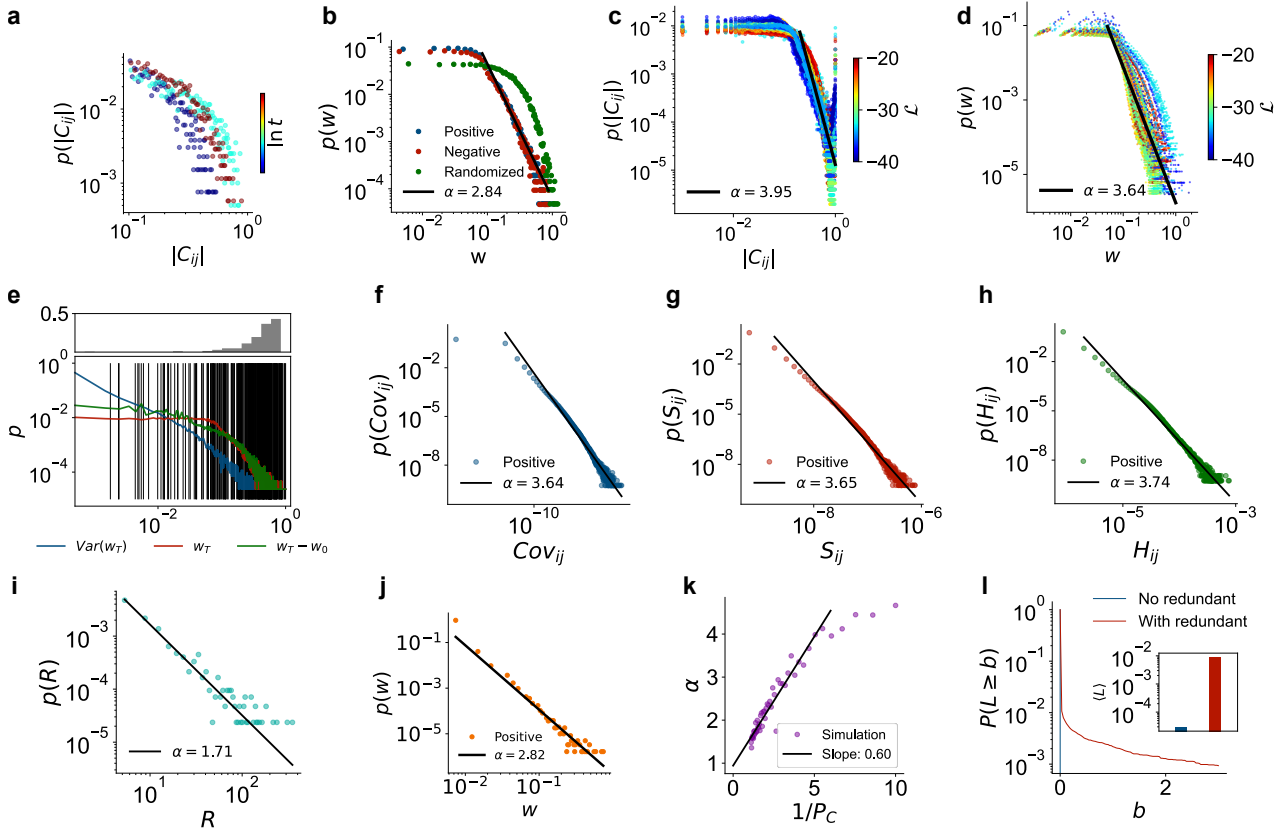}
	\medskip
	\caption{\textbf{Heavy-tailed connectivity and criticality in training.}
	\textbf{a}, Distribution of absolute correlations $p(|C_{ij}|)$ for different training iterations $t=0, 100, 2\times 10^5$. Diagonal entries of the correlation matrix are excluded.
	\textbf{b}, Distribution of positive (blue) and negative (red) connection weights $w$ at the terminal iteration $t=2\times 10^5$, and randomized weights (green). The solid line indicates a power-law fit to the distribution of positive connection weights.
	\textbf{c}, Distribution of correlations at $t = 2 \times 10^5$, for trainings with $\mathcal{L} < -21$. 
	\textbf{d}, Distribution of connection weights $p(w)$ at $t = 2 \times 10^5$, for different hyperparameters. 
	\textbf{e}, Probability distributions of the variance of weight along training trajectory $Var(w_T) = \frac{1}{T} \sum_{t=1}^{T} (w_t - \langle w_t \rangle)^2$, connection weights at the final iteration $w_T$, and connection weight differences between the final and initial iterations $w_T - w_0$. Values for critical connections are indicated with black vertical lines, with the distribution shown in the above sub-figure.  
	\textbf{f-h}, Distribution of gradient noise covariance $C = \text{Cov}(\nabla \hat{L}, \nabla \hat{L}^T)$ (\textbf{f}), average squared gradients $gg^T = \langle \nabla L \rangle_x\langle \nabla L \rangle_x^T$ (\textbf{g}), and Hessian matrix $H = \langle \nabla \nabla L \rangle_x$ (\textbf{h}). 
	The solid black line in each figure represents the power-law distribution fit $p(x) \propto x^{-\alpha}$ to the matrix elements.
	\textbf{i}, Distribution of irreversibility $R = \left\langle \left[Cov(w(t), w(t)) - Cov(w(t), w(-t))\right]^2 \right\rangle_t$ of connection weight along training trajectory. 
	The solid line indicates represents the power-law distribution fit $p(R) \propto R^{-\alpha}$. 
	\textbf{j}, Stable distribution of connection weights (orange) obtained from simulations of the critical redistribution model. Data were accumulated across 100 independent connectivity matrices. The black solid line represents a power-law fit to the weight distribution.
	\textbf{k}, Relation between power-law exponents $\alpha$ and critical probability $P_C$, the solid line indicates a linear relation $\alpha = 1 + 0.6/P_C$. 
	\textbf{l}, Change in the format of concentration inequality with and without redundant connections. The inset sub-figure shows the change in mean loss.  
	}
	\label{fig:criticality}
\end{figure}

The quasi-criticality of NCG and the heavy-tailed distribution of per-sample loss prompt us to investigate the criticality in training ANNs. 
Surprisingly, we found ubiquitous existence of critical behaviors, as shown in Fig.~\ref{fig:criticality}. 
Specifically, the overall correlation (across all tasks) exhibits a heavy-tailed distribution, with intermediate training states ($t=100$) showing even stronger correlations compared to the final iteration (Fig.~\ref{fig:criticality}a), which is consistent with the second-order connectivity phase transition presented in Fig.~\ref{fig:activity_Rastermap}b. 
The connection weights for the converged state also follow a power-law distribution, which clearly deviates from random distribution (Fig.~\ref{fig:criticality}b). 
The heavy-tailed distributions of correlation, connection weight, and per-sample loss are unique to the converged model, whereas unconverged models exhibit dispersed distributions lacking scale-invariance (Fig.~\ref{fig:landscape}c,g; Fig.~\ref{fig:criticality}c,d).

Aside from neuronal connectivity, the dynamical trajectory-related properties all exhibit power-law-like heavy-tailed distributions, such as the variance of trajectory $Var(w)$ (Fig.~\ref{fig:criticality}e), the displacement of weight $w_T - w_0$ (Fig.~\ref{fig:criticality}e), the elements of the covariance matrix of gradient noise $C$ (Fig.~\ref{fig:criticality}f), the average squared gradients $S$ (Fig.~\ref{fig:criticality}g), the elements of the Hessian matrix $H$ (Fig.~\ref{fig:criticality}h), and the irreversibility of per-coordinate trajectory $R$ (Fig.~\ref{fig:criticality}i). 
It is worth noting that $C$, $S$, and $H$ develop heavy-tailed distributions early in training, whereas the connection weights gradually converge to a power-law distribution in the later training stage. 
The coincidence of the second-order connectivity phase transition and the emergence of power-law distributions in gradient-related quantities indicates that the initial training stage corresponds to rapid self-organization of neuronal connections, facilitating multi-scale responses to external changes and thereby supporting adaptive learning.

The neuronal wiring and activity of biological neural circuits often show heavy-tailed statistics, meaning most neurons/regions have few, weak links and low pairwise coupling, while a small minority form hubs and strong pairs. 
This organization likely arises from preferential growth, plasticity, spatial cost constraints, and heterogeneity, enabling efficient communication, robust-yet-vulnerable hub-centric structure, and flexible population dynamics.
Given the heavy-tailed distribution of connectivity and correlations of ANN after training (Fig.~\ref{fig:criticality}a--d), and considering that synaptic density in the biological neural circuits peaks during early childhood before undergoing substantial pruning throughout development \cite{huttenlocherRegionalDifferencesSynaptogenesis1997}, the training behavior of ANNs closely resembles that of biological neural circuits \cite{lynnHeavytailedNeuronalConnectivity2024}, suggesting that both may follow similar learning principles.

To verify this hypothesis, we developed a continuous-valued phenomenological model similar to the Hebbian self-organization in biological neural circuits \cite{lynnHeavytailedNeuronalConnectivity2024}. 
We refer to this as the critical redistribution model, and simulations of this model reproduced the power-law distribution of connection weight with similar exponents for the critical probability $P_C = 0.37$ (Fig.~\ref{fig:criticality}j). 
Remarkable, there is also a linear relationship between the power-law exponent and the critical probability for $P_C > 0.2$ (Fig.~\ref{fig:criticality}k), similarly to results from biological neural circuits \cite{lynnHeavytailedNeuronalConnectivity2024}. 

These results demonstrate that a heavy-tailed distribution of neuronal connectivity can emerge from the correlation structure of ANN activity via a mixed process of random redistribution and targeted reinforcement of critical connections among strongly correlated neurons. 
Accordingly, the model indicates that ANNs exhibit Hebbian-like synaptic plasticity, characterized by a reinforcing positive feedback between neuronal correlation and connectivity. 
SGD implements this process by sampling data to stochastically redistribute connection weights, while backpropagated signals consistently strengthen connections aligned with high neuronal correlations.
Strong connection weights induce high activity correlation during forward propagation, and this elevated correlation yields large gradients during backpropagation, thereby further amplifying the weights.
Consequently, connections between neuron pairs with large correlations are preferentially and persistently enhanced, and they exert greater functional influence than connections among weakly or uncorrelated neurons.

It is important to recognize that the distribution of gradient noise exhibits heavy-tailed behavior, with tails that deviate strongly from Gaussian noise.  
Large deviations in gradient noise correspond to substantial losses for exceptional data samples, which play a critical role in concentrating the loss landscape and should not be neglected.  
Consequently, modeling the stochastic dynamics of SGD in weight space as a Langevin process should also incorporate a power-law-distributed noise covariance as well as high order moments to reproduce the empirically observed power-law distribution of connection weights. 
The corresponding dynamics can deviate strongly from classical Brownian motion and can exhibit anomalous diffusion, aging, and non-Markovian memory effects.

\subsection{Critical directions in loss landscapes}

The critical behavior of neuronal connectivity suggests that a few connections are dominantly significant for the training of ANNs. 
Therefore, we set out to identify critical and non-critical connections using the correlation matrix.
Specifically, for each connection from neuron $i$ to neuron $j$ in the model architecture, if $C_{ij} \ge 0.5$, we classify the weight as critical; otherwise, as non-critical. 

We then examine the shape of the empirical and average landscapes with respect to critical and non-critical connections, as shown in Fig.~\ref{fig:landscape}a. 
In the initial stage, both critical and non-critical connections exhibit large mean loss and landscape divergence. 
Near the connectivity transition point, where NCGs have been extensively established for each target, the landscape along critical directions (corresponding to critical connections) becomes less divergent, in contrast to the non-critical directions. 
In the converged state ($t = 2 \times 10^5$), critical directions exhibit a larger flat region with a higher concentration of loss measure, while non-critical directions remain rugged with a median concentration of loss measure ($t=2\times 10^5$ in Fig.~\ref{fig:landscape}a). 
The training dynamics along critical directions also exhibit slowly decaying autocorrelation and greater irreversibility during the training process compared to non-critical connections (Fig.~\ref{fig:criticality}i), which are signatures of downhill dynamics.

These results suggest that critical connections consistently contribute to the concentration of the empirical landscape and exhibit slow yet irreversible dynamics, whereas non-critical connections contribute little or may even interfere with model performance, displaying rapid fluctuations and quickly reaching an equilibrium-like state.
As a result, the average landscape displays a river-valley \cite{liuNeuralThermodynamicLaws2025} (or non-degenerate-degenerate \cite{yangStochasticGradientDescent2023}) separation ($t=100$ in Fig.~\ref{fig:landscape}a).
More specifically, the gradients along critical directions across time are aligned and exhibit a smooth, flat geometry with consistent drift, indicative of river-like or non-degenerate behavior.
By contrast, the non-critical directions remain divergent in the loss landscape, with sharply varying gradients over small neighborhoods, resulting in highly fluctuating dynamics lacking discernible drift and characteristic of valley-like or degenerate behavior.
Critical connections are also associated with large variance in the weight trajectory, $Var(w_t)$, large final weight $w_T$, and large weight shift $w_T - w_0$ (Fig.~\ref{fig:criticality}e).
This divergence is also evident from the large mean and variance of per-sample gradients.
Together, critical directions favor a smooth, flat loss landscape and also exhibit large trajectory variance, whereas non-critical directions are associated with a choppy, rugged loss landscape and small trajectory variance.
This behavior resembles the inverse variance-flatness relation \cite{fengInverseVarianceFlatness2021}.

Not all critical connections identified near the connectivity transition point survived until the final iteration. 
We term those connections that show high correlation near the connectivity transition point but diminished during later training as redundant connections. 
These redundant connections eventually behave like non-critical connections, exhibiting rapidly decaying autocorrelations and low irreversibility. 
Because their gradients preferentially drive the weights toward opposite signs, they concomitantly depress the absolute magnitudes of these weights. 
Their retention in the terminal state disrupts the concentration of the loss landscape and elevates the average loss (Fig.~\ref{fig:criticality}l). 
Consequently, redundant connections hinder learning and are progressively pruned after the connectivity phase transition.

Alongside the connectivity phase transition in neuronal correlations, these results indicate that ANNs deviate from purely Hebbian plasticity, instead learning by first establishing extensive neuronal correlations and subsequently pruning redundant connections that diverge the loss landscape. 
The emergence of correlated neurons constructs diverse computational graphs linking inputs to predictions, stabilizing outputs and suppressing landscape dispersion. 
In contrast, when inter-layer correlations are absent, downstream neurons yield stochastic, divergent responses, precluding concentration of the empirical landscape. 
Yet overwhelming correlation introduces redundancy and degrades representational capacity. 
Continued optimization after the connectivity transition attenuates such interference by pruning redundant connections.

\setcounter{section}{3}
\setcounter{subsection}{0}
\section{Critical computational graph}
\subsection{Critical computational graph interprets task learning}

The critical neuronal connections indicate that not all neurons are essential for the learning of all digit recognition tasks. 
Notably, the cross-correlations of activities between different tasks are distributed in a degenerate manner, which is further supported by the distinct patterns observed in the Rastermap (Fig.~\ref{fig:activity_Rastermap}a), implying that the ANN has acquired a unique representational signature for each task. 
This clear modularity led us to hypothesize that there exists a distinct core neural graph for each learning task. 
Furthermore, the heavy-tailed distributions of connection weights and neuronal correlations of terminal state suggest that only a few connection weights are significant for the learning task (Fig.~\ref{fig:criticality}a-d). 
Based on these observations, we aim to identify the minimal core neural graph essential for each learning task, which we term the critical computational graph (CCG).

\begin{figure}
	\centering
	\includegraphics[width=\linewidth]{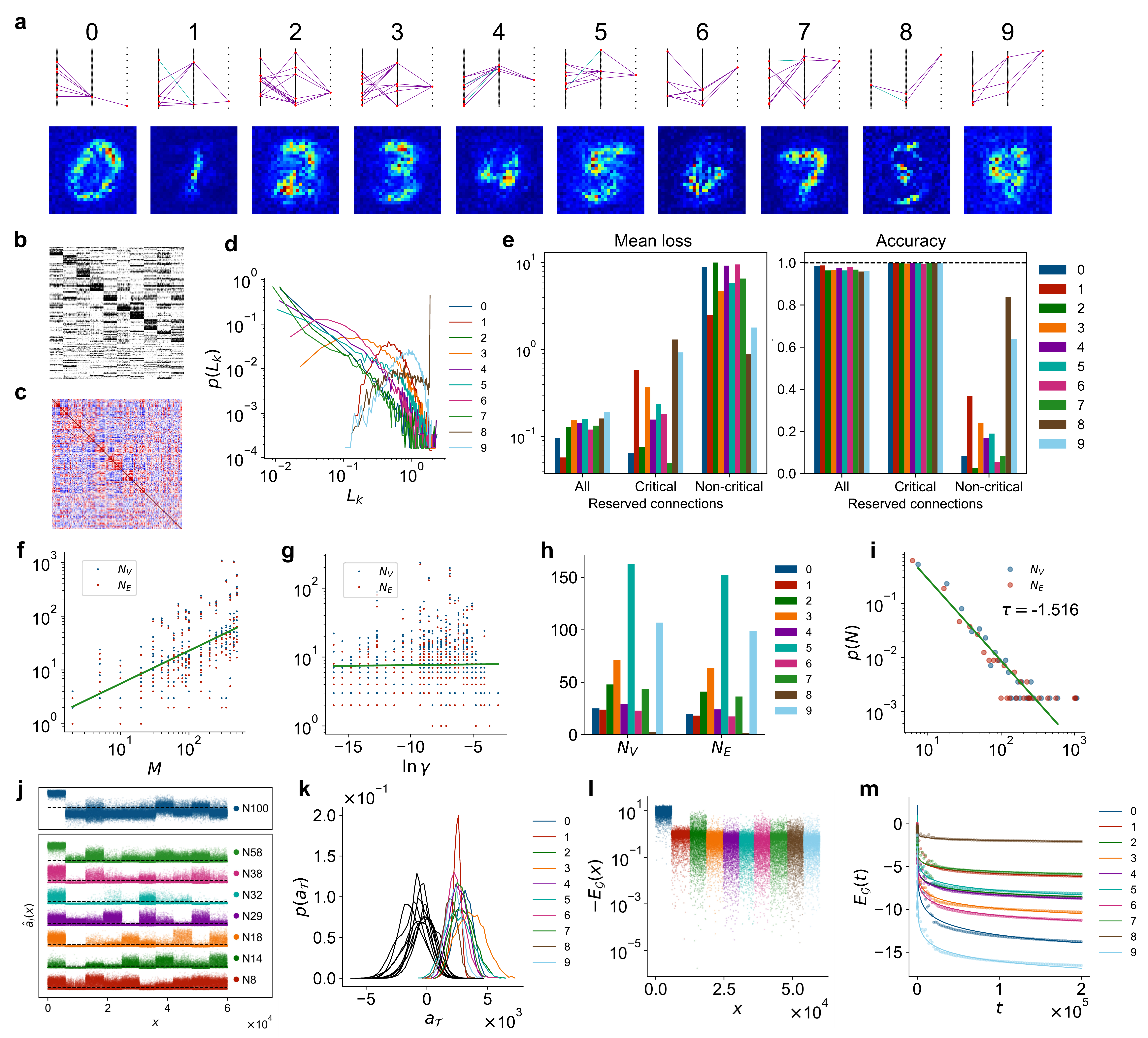}
	\medskip
	\caption{\textbf{Critical computational graph interprets task learning.}
	\textbf{a}, The critical computational graph (CCG) derived using the clustering-growing-pruning strategy for each learning task $\mathcal{T}$ (recognition of digit number $0 \sim 9$), and characteristic features learned by the first layer of each CCG. The colors indicates different neuronal activities. 
	\textbf{b, c}, Rastermap (b) and correlation matrix (c) derived from neurons ordered according to their inclusion in each CCG. 
	\textbf{d}, Distribution of per-samples loss $L_k $ for $x_k \in \mathcal{T}$ of each learning task $\mathcal{T}$. 
	\textbf{e}, Averaged loss and accuracy for each CCG on the test dataset. 
	\textbf{f}, Dependence of the number of critical vertices $N_V$ and critical edges $N_E$ in the CCG on model size $M$. 
	\textbf{g}, Dependence of $N_V$ and $N_E$ in the CCG on the mobility factor $\gamma$. 
	\textbf{h}, Mean number of critical vertices and edges for each learning task. 
	\textbf{i}, Power-law distributions of $N_V$ and $N_E$ obtained by aggregating all learning tasks across all tested hyperparameters.
	\textbf{j}, Coherent coupling of critical neurons in CCG for learning task of digit number 0. 
	\textbf{k}, Activity distribution overlap between task-specific and task-off activities of output neurons.
	\textbf{l}, Per-sample neuronal coupling cost for learning task of digit number 0, color labels are the same as in (b). 
	\textbf{m}, Evolution of neuronal coupling cost for each CCG during training, solid linears represent fitting data points to $E_\mathcal{G} \propto - \alpha \ln t$.
	}
	\label{fig:ccg}
\end{figure}

By employing a clustering-growing-pruning strategy for the terminal state, we are able to identify CCGs from the trained model for each learning task (digital number recognition), as shown in Fig.~\ref{fig:ccg}a. 
Note that these CCGs are much sparser compared to the original densely connected network. 
The average number of edges in these CCGs is on the order of $\mathcal{O}(10)$, whereas in the densely connected network, the number of edges is on the order of $\mathcal{O}(10^4)$, indicating a 1,000-fold reduction in computational cost. 
Furthermore, the edges between each layer pair are reduced by about $\mathcal{O}(10^2)$-fold; for an ANN with $L$ layers, the reduction in computational burden is on the order of $\mathcal{O}(10^{2(L-1)})$. 
Testing the performance of these CCGs on both the training dataset and the test dataset (Fig.~\ref{fig:ccg}e), we found that CCGs produce only a minor increase in average loss, while maintaining adequate accuracy on the training dataset and even achieving higher accuracy on the test dataset.
These results confirm the expectation of Occam's razor: sparse models generalize well, and since CCG is the sparsest model, it generalizes even better than the original model.

To build intuition for CCG, we visualized the primary features obtained by the weighted summation of first-layer neurons for each task (Fig.~\ref{fig:ccg}a). 
Each CCG captures a digit-specific representation by emphasizing distinct pixel subsets. 
These patterns indicate that CCGs render the task-learning dynamics of ANNs interpretable, revealing a hierarchical composition of many weak rules to form predictions. 
Ordering neurons by the CCG sequence for digits 0--9 yields a strikingly structured organization for each task on the Rastermap (Fig.~\ref{fig:ccg}b). 
Correspondingly, essential neuronal covariances concentrate along the diagonal blocks of the correlation matrix (Fig.~\ref{fig:ccg}c). 
Shared connections across tasks are rare, pointing to a modular functional architecture tailored to the tasks, reminiscent of biological neural circuits. 
The per-sample loss from CCG exhibits heavy-tailed behavior for most tasks, while approximating a log-normal distribution for digits 9, 8, and 1 (Fig.~\ref{fig:ccg}d), consistent with the mean loss. 
Notably, the elevated mean loss does not degrade accuracy, implying that the original densely connected network typically overfits the training set. 
By contrast, a sparser model prioritizes the most informative input components and suppresses irrelevant variation, thereby improving generalization.

\subsection{Critical computational graphs are analogous to brain avalanches}


We further investigate how hyperparameters affect CCGs. 
As shown in Fig.~\ref{fig:ccg}f, increasing model size leads to more vertices and edges present in the CCGs of all learning tasks. 
This is consistent with the topology changes of NCG (Fig.~\ref{fig:activity_Rastermap}e), suggesting that larger models rely on many more critical neurons and connections to achieve convergence on the training dataset, and hence are more likely to generalize poorly on the test dataset. 
On the other hand, changing the mobility factor $\gamma$ results in almost the same number of neurons and connections within CCGs (Fig.~\ref{fig:ccg}g), indicating the consistency of representations complexity learned by all models despite training divergence. 
This finding accords with the Platonic representation hypothesis \cite{huhPlatonicRepresentationHypothesis2024}, which posits that learning in ANNs---particularly large language models---rests on universal, rule-based ideal representations. 
Our evidence further indicates that such Platonic representations can be construed as unified encodings of task-specific CCGs shared across models of identical size.
For different model sizes, their learned representations are quite different since the number of vertices and edges in CCG increases with model size (Fig.~\ref{fig:ccg}f).

The complexity of task learning can be measured by the number of vertices and edges in the task-specific CCG. 
As shown in Fig.~\ref{fig:ccg}h, the digits 5 and 9 are the most challenging to acquire, possibly due to the turning and crossing structural patterns. 
Surprisingly, the digit 8 appears as the easiest to learn despite comprising two loops. 
Inspection of the abstract features in the 8-specific CCG reveals that high-valued neuronal activity is concentrated along the upper boundary of the upper loop, the lower boundary of the lower loop, and the central crossing (Fig.~\ref{fig:ccg}a). 
The set of high-valued pixels is minimal, implying that the digit 8 can be encoded with relatively few neurons. 
This counterintuitive result indicates that the trained ANN exploits positional information to identify digits. 
Accordingly, in the absence of data augmentations such as translation, rotation, reflection, or scaling, the ANN preferentially relies on the absolute positions of task-relevant pixels for recognition.

Finally, using CCG, we reveal another remarkable similarity between ANN and biological neural circuits by measuring the distribution of the number of vertices and edges within task-specific CCGs across all learning hyperparameters. 
A clear power-law distribution $P(N) \propto N^{-\tau}$ for vertices and edges is shown in Fig.~\ref{fig:ccg}i, reminiscent of the power-law distribution of avalanche sizes in biological neural circuits. 
Fitting the distribution of vertices (neurons) to a power-law yields $\tau \approx 1.5$, which matches the 3/2 law observed in biological neural circuits across various species (mammals, birds), brain regions, and experimental conditions \cite{Beggs11167, plenzCriticalityNeuralSystems2014}. 
Therefore, task-specific CCGs are analogous to avalanches in the biological neural circuits: the optimal number of connections and neurons for each task follows a power-law distribution with the same exponent as the biological neural circuits.

\subsection{Coherent coupling and logrithmic aging in training}

The emergence of collective neuronal activity behavior during training, as revealed by task-specific CCGs, suggests that dynamic coupling of critical neurons may constitute the foundation for ANN functionality. 
To comprehend this coupling, we track the evolution of neuronal activity across all data samples for each CCG $\mathcal{G}$ for a particular learning task $\mathcal{T} \subseteq \mathcal{D}$ (Fig.~\ref{fig:ccg}j). 
Observable coupling behavior can be identified among these critical neurons, leading to distinguished task-specific activity strength of the output neurons for $x \in \mathcal{T}$ compared to $x \notin \mathcal{T}$. 
However, these couplings are not always constructive, as evident in learning digit 4: neuron 80 contributes negatively compared with positive contributions from neurons 82 and 85. 
Comparing the neuronal activities in learning digit 9, it becomes clear that the function of neuron 80 is to suppress neuronal activities of other neurons for $x \notin \mathcal{T}$. 
Hence, it is not merely correlation, but essentially cooperative neuronal contributions, that leads to the minimal distribution overlap of neuronal activities between $x \in \mathcal{T}$ and $x \notin \mathcal{T}$ (Fig.~\ref{fig:ccg}k). 
This minimized overlap of activity distribution is consistent with the reduction of prediction entropy in Fig.~\ref{fig:activity_Rastermap}d.

We term this phenomenon, whereby strongly activated neurons function in a data-coordinated manner to cooperatively contribute task-specific, consistent, and distinguished signals to next-layer neurons, as \emph{coherent coupling of critical neurons}. 
This coherent coupling can be quantified using the coupling cost $E_\mathcal{G}$, analogous to spin glass models. 
Low coupling cost represent high coherent coupling, similar to ``energy''. 
Note that coherent coupling extends beyond mere correlation, as it also requires a large signal-to-noise ratio, meaning that coherent coupling represents the coupling of activity differences between $x \in \mathcal{T}$ and $x \notin \mathcal{T}$. 
Therefore, reducing the coupling cost results in attraction of neuronal activities for $x \in \mathcal{T}$, while simultaneously producing repulsion of neuronal activities for $x \notin \mathcal{T}$. 
In the digit classification problem, the cross-entropy loss function is applied after softmax and is minimized when there exists maximal neuronal activity divergence between task-specific and task-irrelevant neuronal activities. 
For precise value-based loss functions, we additionally expect the divergence of neuronal activities to be minimized. 
In this case, we can incorporate the activity variance into the coupling cost.

At the end of training, per-sample coupling for $x \in \mathcal{T}$ exhibits distinct coupling cost compared to $x \notin \mathcal{T}$ (Fig.~\ref{fig:ccg}l). 
The net result is a significant reduction in the coupling cost for neurons in the CCG $\mathcal{G}$ for a specific learning task $\mathcal{T}$. 
Note that for some $\mathcal{T}$ (digits 0, 1, 2, 3, 5, 7), coherent coupling exists even in the initial state; hence, training corresponds to a gradual reinforcement of these couplings. 
For other $\mathcal{T}$ (digits 4, 6, 8, 9), coherent coupling gradually emerges during training in the very early training stages $t \in (1, 100)$. 
Therefore, the connectivity phase transition observed in Fig.~\ref{fig:activity_Rastermap}e-g and Fig.~\ref{fig:landscape}f corresponds to the establishment of coherent coupling in the early stage and gradual refinement in the later stage.

Throughout the entire training process, we observe a monotonic reduction in the coupling cost, exhibiting slow relaxation dynamics characterized by a logarithmic time dependence $E_\mathcal{G}(t) \propto -\ln t$ (Fig.~\ref{fig:ccg}m), typically observed in the aging of glassy or disordered systems \cite{berthierStatisticalMechanicsPerspective2009}. 
We provide a simple explanation of the logarithmic aging phenomenon based on the criticality of training. 
From this perspective, training ANNs is analogous to the simultaneous relaxation of many entangled disordered systems, each corresponding to a learning task.
This behavior is consistent with the decrease in prediction entropy (Fig.~\ref{fig:activity_Rastermap}d), the convergence of connection weight distribution to a power-law distribution (Fig.~\ref{fig:landscape}c,g), and the increase in quasi-thermal entropy $\Delta S$ (Fig.~\ref{fig:landscape}d). 
These collective behaviors represent the strong interplay between neuronal activities, connection weights, and the empirical loss landscape. 

\setcounter{section}{4}
\setcounter{subsection}{0}
\section{Learning through self-organized assembly of coherently coupled critical neurons}

\begin{figure}
	\centering
	\includegraphics[width=0.8\linewidth]{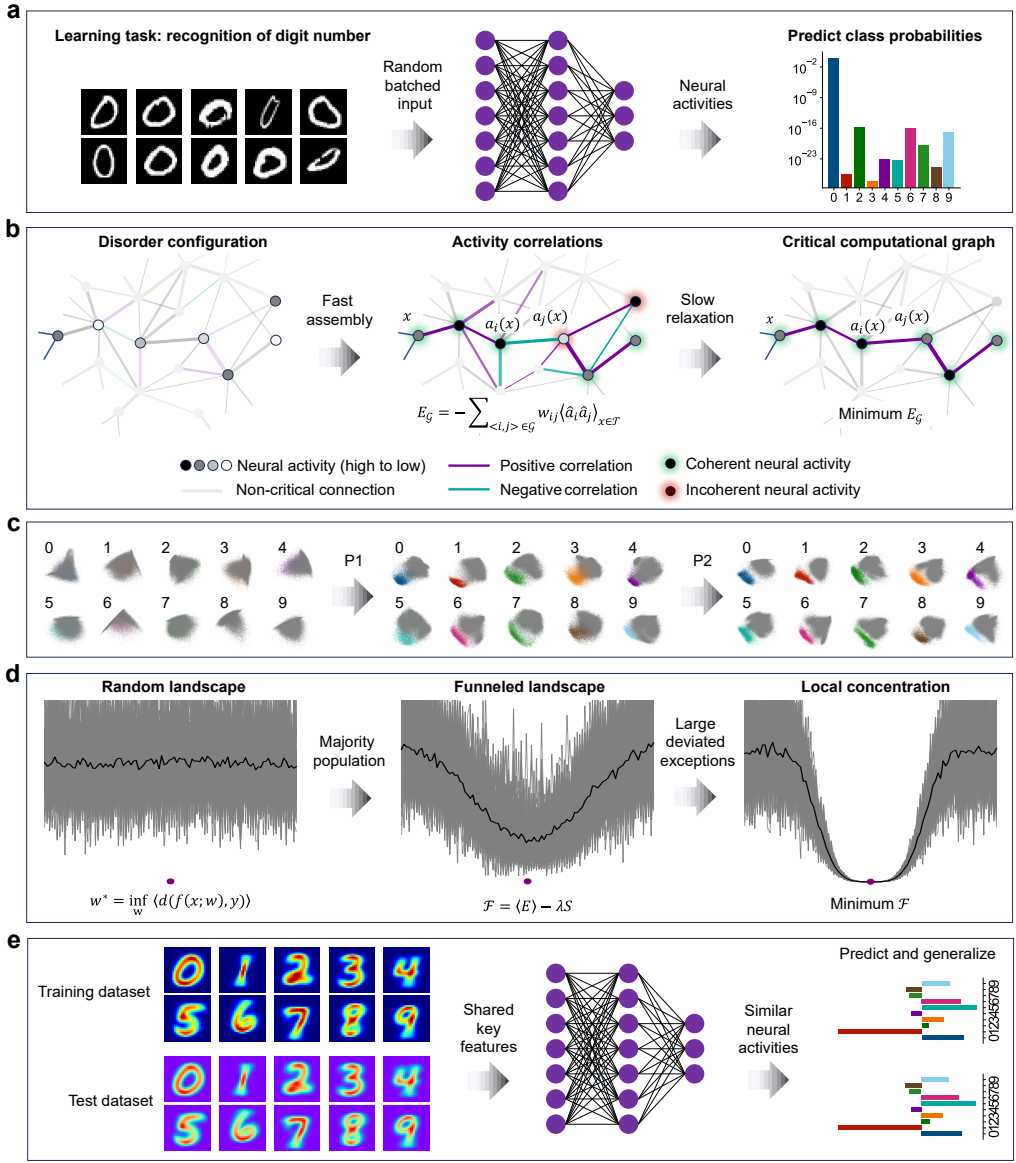}
	\medskip
	\caption{\textbf{Neural network learns via coherent coupling of critical neurons.}
	\textbf{a}, Scheme of learning process of ANN. 
	The digit classification task is addressed by iteratively updating ANN connection weights via backpropagation, minimizing the divergence between randomly batched inputs and the predicted class probabilities. 
	\textbf{b}, Deep artificial neural networks acquire computational capabilities through the self-organization of coherently coupled critical neurons, forming critical computational graphs.
	In the initial stage, SGD rapidly induces correlations among neurons and assembles them into computational graphs, albeit with incoherent task-specific neuronal activities.
	Subsequently, a slow relaxation of coupling cost between critical neurons enables the system to achieve coherent neuronal activity coupling tailored to specific tasks, leading to the establishment of critical computational graphs.
	\textbf{c}, Coherent coupling among critical neurons induces a geometric deformation of the critical neural manifold, thereby segregating task-specific neuronal activity (colored dots). P1 denotes an energy-driven assembly phase of correlation structure associated with a connectivity phase transition, whereas P2 denotes an entropy-driven phase of precise neuronal activity coupling, characterized by a logarithmic aging process culminating in convergence as quantified by the convergence score.
	\textbf{d}, Coherent coupling of critical neurons is equivalent to the local concentration of loss landscapes (gray dashed lines), as compared to the optimal connection weight $w^*$ (represented as a small purple dot).
	\textbf{e}, Training and test datasets share common key features that engage the same critical computational graphs, eliciting similar neuronal activity patterns and thereby extending predictive performance to previously unseen samples. 
	}
	\label{fig:framework}
\end{figure}

To formulate a unified self-consistent description, we recognize that ANNs function via hierarchical abstraction of features present in the dataset. 
According to the manifold hypothesis   natural data often exhibit fractal-like structures, a finite set of key features---latent variables, or reaction coordinates---is typically sufficient to capture the essential patterns relevant to a specific learning task. 
As evident from the performance of CCGs (Fig.~\ref{fig:ccg}e), critical neurons linked by critical connections exactly captures these task-specific key features and predominantly determine the generalized predictive capability of ANNs. 
These critical neurons are highly activated and coherently coupled, as evidenced in Fig.~\ref{fig:ccg}j.
Their population activity during a learning task lies on a low-dimensional, smooth neuronal manifold, embedded in the high-dimensional activity space of all neurons (Fig.~\ref{fig:framework}c) \cite{perichNeuralManifoldView2025}.
In contrast, irrelevant noisy details in the data, as well as task-irrelevant and non-critical neuronal activities, contribute randomly fluctuating signals to neurons in subsequent layers. 

Since the dimensionality of task-relevant key features is significantly lower than that of the input space, optimal performance requires that the ANN develop a representation whose intrinsic dimensionality aligns with that of the key features while attenuating dependence on irrelevant information. 
This alignment is enforced by minimizing the coupling cost among critical neurons. 
Reducing coupling cost yields persistent reinforcement of critical connections and induces a heavy-tailed distribution of connection weights (Fig.~\ref{fig:criticality}b,d). 
Because the SGD gradient scales with the product of weights along the backpropagation path, its fluctuations are likewise heavy-tailed, leading to power-law distributions of the trajectory variance, the Hessian spectrum, and associated metrics (Fig.~\ref{fig:criticality}e--i). 
Under a first-order approximation in which neuronal correlations scale with connection strength, the weight heavy tails promote heavy-tailed neuronal correlations. 
The resulting effective network is characterized by sparse, critical connections among critical neurons (Fig.~\ref{fig:criticality}a--d; Fig.~\ref{fig:ccg}a,i), generating a scale-free correlation structure with multiscale responses that align with the dimensionality of key features while suppressing irrelevant input noise and task-irrelevant, non-critical activity.

Through this mechanism, direct statistical dependencies can be established among critical neurons, thereby linking input neurons to output neurons via associative computation pathways. 
Enhancing the pairwise correlations among these critical neurons can reduce the divergence of loss landscapes, enabling precise computation without explicit logical operations. 
Therefore, minimizing the coupling cost of critical neurons effectively leads to the concentration of measure in loss landscapes. 
In conclusion, the coherent coupling of critical neurons offers a unified theoretical framework explaining the emergence of correlation patterns in neuronal activities, the local concentration of loss landscapes, and criticality in training dynamics.

The entire training process can be comprehensively understood within our framework.
In the initial training stage (Fig.~\ref{fig:framework}b, left panel), connections with weak neuronal correlations are prevalent, while the distribution of critical connections remains largely random. 
Following initial training iterations, NCGs emerge spontaneously through SGD induction (Fig.~\ref{fig:framework}b, middle panel; Fig.~\ref{fig:activity_Rastermap}a), analogous to modules formed in self-organizing systems. 
During subsequent training phases (Fig.~\ref{fig:framework}b, right panel), refinement and pruning of connection weights deform the neural manifold to minimize activity overlap (Fig.~\ref{fig:framework}c, right panel), thereby lowering the coupling cost and enabling precise, coherent coupling of critical neurons in CCG. 
The converged model corresponds to the one with minimum coupling score (Fig.~\ref{fig:ccg}m). 
This adjustment of connection strengths to reduce the coupling score parallels protein evolution, wherein interaction energies among amino acids are tuned to minimize frustration and maximize stability \cite{bryngelsonSpinGlassesStatistical1987}.
This progressive coherent coupling of neuronal activity ensures robust and predictable output along established associative pathways.

The behavior of empirical loss landscapes directly reflects the collective behavior of neuronal activities. 
In the initial phase (Fig.~\ref{fig:framework}c, left panel), the empirical landscape $\hat{L}$ is characterized by an effectively uncorrelated random field in the vicinity of the optimal connection weight vector $w^*$, whose existence is guaranteed by the universal approximation theorem. 
Due to randomly distributed critical connections, the ensemble-averaged loss landscape $\bar{L}$ is generally rough (Fig.~\ref{fig:landscape}a, left panel), ostensibly rendering convergence to a stable model intractable. 
As the neuronal correlation structure consolidates (Fig.~\ref{fig:framework}c, middle panel), a decrease in overall loss occurs for the majority population of data, reflected in the emergence of spatial correlation near $w^*$ in the empirical loss landscape, manifesting as a funneled landscape geometry. 
Finally (Fig.~\ref{fig:framework}c, right panel), further coherent coupling of neuronal activities, corresponding to targeted optimization of connection weights for exceptions with large losses, reduces the roughness of the empirical loss landscape. 
This results in concentration of measure within a local subspace across empirical loss landscapes, quantitatively described by the free energy $\mathcal{F}$.
Given that training and test samples typically share key features, the coherent coupling of neuronal activity, together with the ensuing concentration of the loss landscape, yields similar activity patterns, thereby enabling precise prediction and robust generalization to unseen samples (Fig.~\ref{fig:framework}e).

\section*{Discussions}


Conventional optimization theory focuses solely on reducing the global loss, viewing optimization as the movement of the model state along the mean loss landscape $\bar{L}$ in an attempt to find the global minimum of $\bar{L}$. 
In this perspective, the convergence of training in over-parameterized ANNs using SGD presents both a practical certainty and a theoretical enigma: despite the highly non-convex and rugged nature of their loss landscapes , over-parameterized ANNs often achieve remarkably strong empirical performance. 
Nonetheless, key open questions remain regarding the statistical and geometric principles that enable standard optimization algorithms, such as SGD, to successfully escape saddle points and locate local minima with both low training and test errors in these high-dimensional landscapes. 
Although training is remarkably robust to variations in initialization, architecture, learning rate, and other hyperparameters, there remains a limited theoretical understanding of the mechanisms underlying such robustness and rapid convergence, especially since classical optimization guarantees typically do not apply in this context. 
Until recently, no general global convergence guarantees were known for the training of deep, nonlinear ANNs, and most existing results either rely on unrealistic assumptions or address only restricted cases. 
In fact, during the training process of ANNs using SGD, it is impossible to navigate the whole rugged empirical loss landscape (Fig.~\ref{fig:landscape}a). 
The change in the weight vector is also very limited, as compared to the region of large mean loss ($t=1$ in Fig.~\ref{fig:landscape}a), suggesting that the converged state is very close to the initial state.

Our findings reveal a distinct mechanism to explain this dilemma. 
As illustrated in Fig.~\ref{fig:framework}, the training of ANNs should no longer be regarded as a deterministic gradient descendant process that solely seeks a global minimum. 
Instead, it is better conceptualized as an evolutionary process within a self-organizing system, driven by the formation of CCGs through coherent coupling of task-specific critical neurons. 
Consequently, the empirical loss landscape exhibits a locally concentrated measure with vanishing divergence.
In tandem, minimal loss-landscape divergence, together with the universal approximation theorem, provides a theoretical guarantee of the convergence of ANN training, even in the presence of random data and stochastic computation.
Assuming that test samples share key features with the training distribution, the resulting predictive performance extends to previously unseen data.
This phenomenon aligns with demands from real-world applications, such as autonomous driving, robotics, and conversational AI, where trained ANNs must ensure not only accurate predictions but also consistent reliability to prevent one-off mistakes or hallucinations. 
Specifically, SGD works to reduce prediction variance by strengthening the connectivity among correlated neurons during training, following a pattern similar to the critical redistribution manner. 
Furthermore, locally concentrated regions correspond to domains that are crucial for ensuring the stability of solutions identified by SGD. 
The sparsity of critical connections in ANN training offers an alternative perspective on the saddle-point problem. 
Because only saddle points aligned with critical directions are consequential, whereas those along non-critical directions are negligible, the number of effective saddle points does not scale combinatorially. 
This sparsification therefore potentially explaining why training can be so efficient.


The formation of NCGs during coherent coupling closely resembles Hebbian plasticity in biological neural circuit learning, wherein synaptic connections evolve in response to correlations in neuronal activities \cite{mageeSynapticPlasticityForms2020}. 
However, we observed a second-order phase transition in both neuronal correlation strength and the topology of NCGs, a phenomenon that cannot be explained by Hebbian plasticity alone. 
Beyond neuronal correlations, coherent coupling necessitates the alignment of neuronal activities with high signal-to-noise ratios across different data samples sharing common features, while simultaneously discouraging the co-activation of neurons in response to data samples with divergent features. 
Such a mechanism promotes the elimination of multifunctional redundant connections and supports the formation of task-specific, noise-isolated, associative computational pathways with minimal cross-talk. 
This process accounts for the observed flourish-diminish dynamics of neuronal correlations during training, which mirrors patterns observed in most biological neural circuit development \cite{tianTheoreticalFoundationsStudying2022}. 
The resulting modularity represents a desirable property for neural computation in both artificial and biological neural networks.

Another remarkable parallelism is the identical power-law distributions in CCG size and avalanche size, which also accounts for the similar heavy-tailed distributions in connection weights and neuronal correlations. 
Therefore, it is reasonable to hypothesize that the nature of avalanches in biological neural circuits might result from CCG activation triggered by input stimuli under evolution requirement. 
This argument suggests that the brain functions using a minimal task-specific neural network, which has considerably fewer neurons and connections than a densely connected network. 
Consequently, the brain is highly energy-efficient compared to ANNs, yet still performs at the same level or even better, since sparse representations almost always generalize well. 
The attempts to maximize generalization in the brain also suggest that the number of neurons in the brain must be controlled; that is, an optimal brain size, rather than a very large brain size, is suitable for survival. 
The mismatch between brain size and the complexity of environmental stimuli may explain why general-purpose intelligence is so rare.
For example, in niche narrowing, the evolutionary shift from a broad ecological niche---using many resources, habitats, or strategies---to a narrower one with fewer, more specialized resources or behaviors.
The species can become extremely efficient in a limited set of conditions, but often at the cost of flexibility and general-purpose cognitive complexity.

ANNs and biological neural circuits share substantial similarities; for instance, both effectively perform a wide range of cognitive tasks and demonstrate comparable functional, structural, and dynamical properties \cite{fristonFreeenergyPrincipleUnified2010, lillicrapBackpropagationBrain2020, richardsDeepLearningFramework2019}. 
However, no mechanistic analogy has been identified between artificial and biological neural networks. 
Our findings indicate that ANNs and biological neural circuits operate in more analogous ways than previously recognized by the scientific communities in both artificial intelligence and neuroscience. 
This striking parallelism further suggests that coherent coupling of critical neurons might represent a potentially unified learning mechanism for both artificial and biological neural networks. 
However, the approaches to achieving predictive capability may differ: ANNs train through a global update strategy, wherein all weight parameters get updated for every training iteration based on gradients of the loss function.
On the contrary, biological neural circuits may utilize a local strategy based on local cellular information to directly reinforce coherent coupling to achieve generalized uniform performance. 
These two approaches thus represent two sides of the same coin.
Therefore, it is compelling to investigate whether these coherent coupling phenomena of critical neurons similarly manifest in biological neural systems, and how CCGs emerge through development.


The self-organized NCGs in early-stage training undergo a connectivity phase transition in complex networks, wherein a giant component emerges through SGD induction. 
The final converged network represents a quasi-critical state, characterized by a scale-invariant distribution of connection degrees and clique sizes. 
This behavior resembles self-organized quasi-criticality \cite{tianTheoreticalFoundationsStudying2022}, where the terminal model ($t = 2 \times 10^5$) remains continually at the ``edge of chaos,'' enabling maximal computational and functional capabilities with efficient communication and long-range correlations, which are essential for ANN functionality. 
However, as model size increases, the connectivity phase transition vanishes, indicating that this phase transition results from the limited number of neurons, i.e., the cross-talk among finite neurons. 
Large models tend to utilize additional neurons to gradually correct predictions rather than attempting to obtain correct abstractions of data features. 
However, these detailed corrections are data-sensitive, thus explaining why large models exhibit poor generalization. 
Understanding and manipulating the cross-talk among neurons during training will reveal additional details of the connectivity phase transition and potentially enhance generalization, which may also offer insights into the quasi-criticality of the human brain.

A deep analogy can be drawn between our proposed minimum quasi-thermal free energy principle and thermodynamic principles, wherein training ANNs is analogous to the process of approaching equilibrium in thermodynamic systems.
The hierarchical abstraction processes inherent in ANN computation closely resemble a gradual coarse-graining procedure, akin to the thermalization observed in natural dynamical systems. 
The concentration of empirical landscapes around specific error values is analogous to the concentration of macro-variables over time as a system approaches equilibrium in thermodynamics. 
However, in a predefined thermodynamic system, the properties of the equilibrium state are strictly determined by thermodynamic laws. 
In contrast, for ANNs, the concentration of empirical landscapes is governed by both the training dataset and the specific form of the loss function, indicating that ANNs can be viewed as purpose driven thermodynamic systems. 
It is therefore compelling to examine whether, and in what manner, a physical thermodynamic system can be repurposed as an intelligent system, and to delineate the conditions under which such intelligence transition can occur.

To quantify the coherent coupling of critical neurons, we propose the coupling cost $E_\mathcal{G}$, which resembles the form of spin glass and Hopfield networks \cite{hopfieldNeuralNetworksPhysical1982}. 
However, $E_\mathcal{G}$ is computed for a specific computational graph corresponding to a particular learning task, with $\mathcal{G}$ as a subgraph of the complete neural network. 
In contrast, the energy functional for Hopfield networks is defined for the entire neural network and for a sequence of storage states belonging to a single learning task. 
Furthermore, the definition of $E_\mathcal{G}$ utilizes the signal-to-noise ratio of the neuronal activities, which provides a measure of the activity differences between data samples belonging to a learning task and other data samples, effectively enhancing the repulsion of predictions for data samples with divergent features.
Given the discrete-valued spins in the Hopfield network, the coupling cost follows a non-increasing staircase with unpredictable step heights. 
Meanwhile, $E_\mathcal{G}$ decreases monotonically, exhibiting a logarithmic time dependence analogous to the physical aging process observed in many disordered systems.
Finally, minimizing the coupling cost yields a critical computational graph, indicating that, after training, the ANN memorizes a specific functional form tailored to data with particular features. 
In contrast, Hopfield networks memorize discrete states rather than functions. 
Consequently, Hopfield networks exhibit an intrinsic upper bound on the number of storable samples, whereas ANNs leverage a learned function to generalize across potentially unbounded sets of similarly structured inputs. 
Thus, our findings highlight a sharp distinction between ANNs and Hopfield networks in their underlying computational mechanisms.


In summary, the coherent coupling of critical neurons provides a comprehensive theoretical framework for elucidating the training dynamics of neural networks, while simultaneously revealing fundamental connections among artificial neural networks, biological neural circuits, and thermodynamic processes in nature. 
The proposed mechanism will facilitate the development of energy-efficient training algorithms and has the potential to advance the engineering of brain-inspired intelligent systems.
The present study focuses primarily on shallow, fully connected artificial neural networks trained on the MNIST dataset. 
Although the scope is restricted to these conditions, we anticipate that the core principles and findings reported here will extend to more complex datasets and larger models with varied architectures, albeit with potentially richer and more sophisticated dynamical behaviors.

\setcounter{section}{0}
\section*{Methods}
Detailed methods and supporting information will be made available upon formal publication.

\clearpage

\section*{Data \& code availability}
The data that support the findings of this study are available from the corresponding authors upon reasonable request.

The codes used in this study are available from the corresponding authors upon reasonable request.


\section*{Acknowledgments}
C.L. thanks the supports from the National Natural Science Foundation of China Grant 32000888, the Scientific Instrument Developing Project of the Chinese Academy of Sciences Grant YJKYYQ20180038,
Jilin Province Science and Technology Development Plan Grant 20230101152JC.

\section*{Author contributions}
J.W. and C.L. designed the study, developed theoretical concepts and tools, analyzed the data, and wrote the paper. 
C.L. developed the analyzing algorithms, performed the numerical experiments, and developed methodology tools to interpret the results.

\section*{Competing interests}
The authors declare no competing interests.

\section*{Correspondence}
Correspondence and requests for materials should be addressed to Jin Wang.

\section*{Additional information}
\textbf{Supplementary information} is provided in the last part of this file. 

\bibliographystyle{naturemag}

\clearpage

\end{document}